\documentclass[journal=jacsat,manuscript=article]{achemso}
\setkeys{acs}{articletitle=true}

\usepackage[version=3]{mhchem} 
\usepackage[T1]{fontenc}       
\usepackage{array}

\usepackage{amsmath,amssymb}
\usepackage{graphicx,color}
\usepackage{subcaption}
\usepackage{makecell}
\usepackage{comment}
\usepackage{etoolbox}

\makeatletter
\patchcmd{\ttlh@hang}{\parindent\z@}{\parindent\z@\leavevmode}{}{}
\patchcmd{\ttlh@hang}{\noindent}{}{}{}
\makeatother

\usepackage{xr}
\newcommand*{\dd}{\mathrm d}
\newcommand*{\ww}{\mathrm w}
\newcommand*{\hh}{\mathrm h}
\newcommand*{\aaa}{\mathrm a}
\newcommand*{\mm}{\mathrm {max}}

\newcommand*{\sGL}{\sigma_{\scriptscriptstyle\rm GL}}
\newcommand*{\sSL}{\sigma_{\scriptscriptstyle\rm SL}}
\newcommand*{\sSG}{\sigma_{\scriptscriptstyle\rm SG}}

\newcommand*{\ttc}{\theta_{\scriptscriptstyle\rm C}}
\newcommand*{\tty}{\theta_{\scriptscriptstyle\rm Y}}

\author{Marion Silvestrini}
\author{Carolina Brito}
\email{carolina.brito@ufrgs.br}
\affiliation[Universidade Federal do Rio Grande do Sul]
{Instituto de F{\'\i}sica, Universidade Federal do Rio Grande do Sul. Av. Bento Gon\c{c}alves 9500, CEP 91501-970, Porto Alegre/RS -  Brazil}

\title[reentrant surfaces]
{Wettability of  reentrant surfaces: a global energy approach}


\begin{document}


 \begin{abstract}
In this work we consider two possible wetting states for a droplet when  placed on a substrate: the Fakir configuration of a Cassie-Baxter (CB) state with a droplet residing on top of roughness grooves and the one characterized by the homogeneous wetting of the surface, referred as the Wenzel (W) state. 
We extend a theoretical model based on the global interfacial energies for both states CB and W to study the wetting behavior of simple and double reentrant surfaces. 
Due to the minimization of the energies associated to each wetting state, we predict the thermodynamic wetting state of the droplet for a given surface texture and obtain its contact angle $\ttc$. We first use this model to find the geometries for pillared, simple and double reentrant surfaces that most enhances $\ttc$ and conclude that the repellent behavior of these surfaces is governed by the relation between the height and width of the reentrances. 
We compare our results with recent experiments and discuss the limitations of this thermodynamic approach. 
To address one of these limitations, we implement Monte Carlo simulations of the cellular Potts Model in three dimensions, which allow us to investigate the dependency of the wetting state on the initial state of the droplet.
We find that when the droplet is initialized in a CB state, it gets trapped in a local minimum and stays in the repellent behavior irrespective of the theoretical prediction. 
When the initial state is  W, simulations show a good agreement with theory for pillared surfaces for all geometries, but for reentrant surfaces the agreement only happens in few cases: for most simulated geometries  the contact angle reached by the droplet in simulations is higher than  $\ttc$ predicted by the model. Moreover, we find that the contact angle of the simulated droplet is higher when placed on the reentrant surfaces than for a pillared surfaces with the same height, width and pillar distance. 
\end{abstract}

\section*{Introduction}

Understanding the parameters that control the wetting properties of a substrate is important to engineer surfaces with different applications.  
One of the ingredients to control the wetting phenomenology is the chemistry of the surface as well as the chemistry of the fluid.
For an idealized surface completely flat, the droplet contact angle is univocally defined by minimizing the necessary energies to generate the interfaces of the three involved phases: it defines the Young contact angle $\tty$, which depends on the surface tension between the liquid-gas $\sGL$, gas-solid $\sSG$ and solid-liquid $\sSL$,  $\cos(\tty) = (\sSG -\sSL) / \sGL$. Another controlling parameter  is the topology of the substrate.
To transform materials for which $\tty > 90^{\circ}$ into a  super-repellent surface  (usually defined as a surface for which the aparent contact angle of a drop of liquid deposited on it is typically $> 150^{\circ}$ and the hysteresis effect is small) is possible by introducing roughness on multiple scales \cite{Quere2008}. This mechanism is understood due to the inspiration in the natural surfaces as the Lotus leaves and to numerous experiments, models and simulations\cite{Nosonovsky2007,Li2007, Quere2008,shirtcliffe2010,Ramos2009, deGennes1985,Patankar2003,Marmur2003,Koishi2009,Koishi2011,Giacomello2012,Shahraz2013,Lopes2013, Patankar_Langmuir2010}.

While the super-repellent behavior for materials and liquids with $\tty > 90^{\circ}$ can be explained by the complementary roles of surface energy and roughness, in the case where $\tty < 90^{\circ}$ the understanding requires more elements.
In the reference \cite{Baumberg2005} the authors have demonstrated that gold surfaces which is hydrophilic with a contact angle of $70^{\circ}$ for water  became hydrophobic (contact angles of the water droplet $ > 90^{\circ}$) when decorated with spherical cavities. This behavior was theoretically discussed by Pantakar \cite{Patankar2009}. 
A superoleophobic surface was also possible from an intrinsically oleophilic (contact angles of the oil droplet $ < 90^{\circ}$) material by building  a hierarchical porous structure consisting of micrometer-sized asperities superimposed onto a network of nanometer-sized pores  \cite{cao2008super}.
In \cite{TutejaSicence2007, TutejaPNAS2008} super-repellent surfaces were developed for organic liquids having  lower surface tensions than that of water. 
Although the thermodynamics of these surfaces show that the global minimum energy state of a droplet placed on this surface would be wetted, the authors have shown that it is possible to design metastable super-repellent surfaces even with materials with $\tty < 90^{\circ}$ and that to understand this behavior the reentrant surface local curvature is determinant.
Other reentrant surfaces with super-repellent  properties for liquids with varying surface tension liquids \cite{park2016analysis} were developed and recently Liu and Kim show that a specific double reentrant structure can render the surface of any material super-repellent \cite{Liu2014},  even for liquids with extremely low surface tension.
It is important to note that the  presence of reentrant curvature is not a sufficient condition for developing highly non-wetting surfaces. Using a free energy model  combined to a hydrodynamic equation, it was shown that reentrant geometries can provide metastable super-repellent states even when the surface is intrinsically wetting  \cite{joly2009}. Also some simulations were developed to measure the energy barrier between the  super-repellent  and wetting states \cite{Escobedo2012a, Escobedo2012b} and to test the robustness of the superomniphobic behavior \cite{Zhang2015, brown2015mechanically}, as well as experiments to better understand its properties \cite{chen2015, Ramos2016}.

Inspired by Kim's experiment \cite{Liu2014}, in this work we extend a theoretical model developed for pillared surfaces in the reference \cite{Fernandes2015} to the surfaces with a simple and double reentrance, as  schematized in Fig.(\ref{typesSurface}). 
The theoretical continuous model takes into account all the interfacial energies associated to the energy of a liquid droplet deposited on top of the surfaces. We consider the Fakir Cassie-Baxter state (CB),  characterized by the suspension of the droplet trapping air inside the surface grooves, and the Wenzel state (W),  where the liquid present a homogeneous wetting of the surface. To obtain the stable wetting state, the energies associated with W and CB states are minimized. This model and the minimization procedure  allow us to build the wetting diagram of the three types of  surfaces for different geometric parameters and types of liquids. We compare the results of this approach with some experiments and discuss the limitations of the model. As mentioned above, a relevant aspect of the wetting problem is the metastability of the wetting states. This metastability in some experiments is manifested through the dependence of the final wetting state of the droplet on its initial condition~\cite{Quere2008, Koishi2009}. To address this issue we implement  Monte Carlo (MC) cellular Potts model simulations~\cite{Graner1992,  Lopes2013}  of a droplet in three dimensions. The simulations show that when the initial wetting state of the droplet is CB, the droplet stays in a non-wetting state during all the simulation run and it usually  reaches a {\it local} minimum. For pillared surfaces, the simulations have good agreement with the theoretical model when the droplet starts in a W configuration, but for reentrant surfaces the simulated angle is always higher than the model predicts. 

\section*{The continuous model}

\begin{figure}
\begin{minipage}{.33\textwidth}
  \begin{subfigure}{\linewidth}
    \centering
    \includegraphics[width=.95\linewidth]{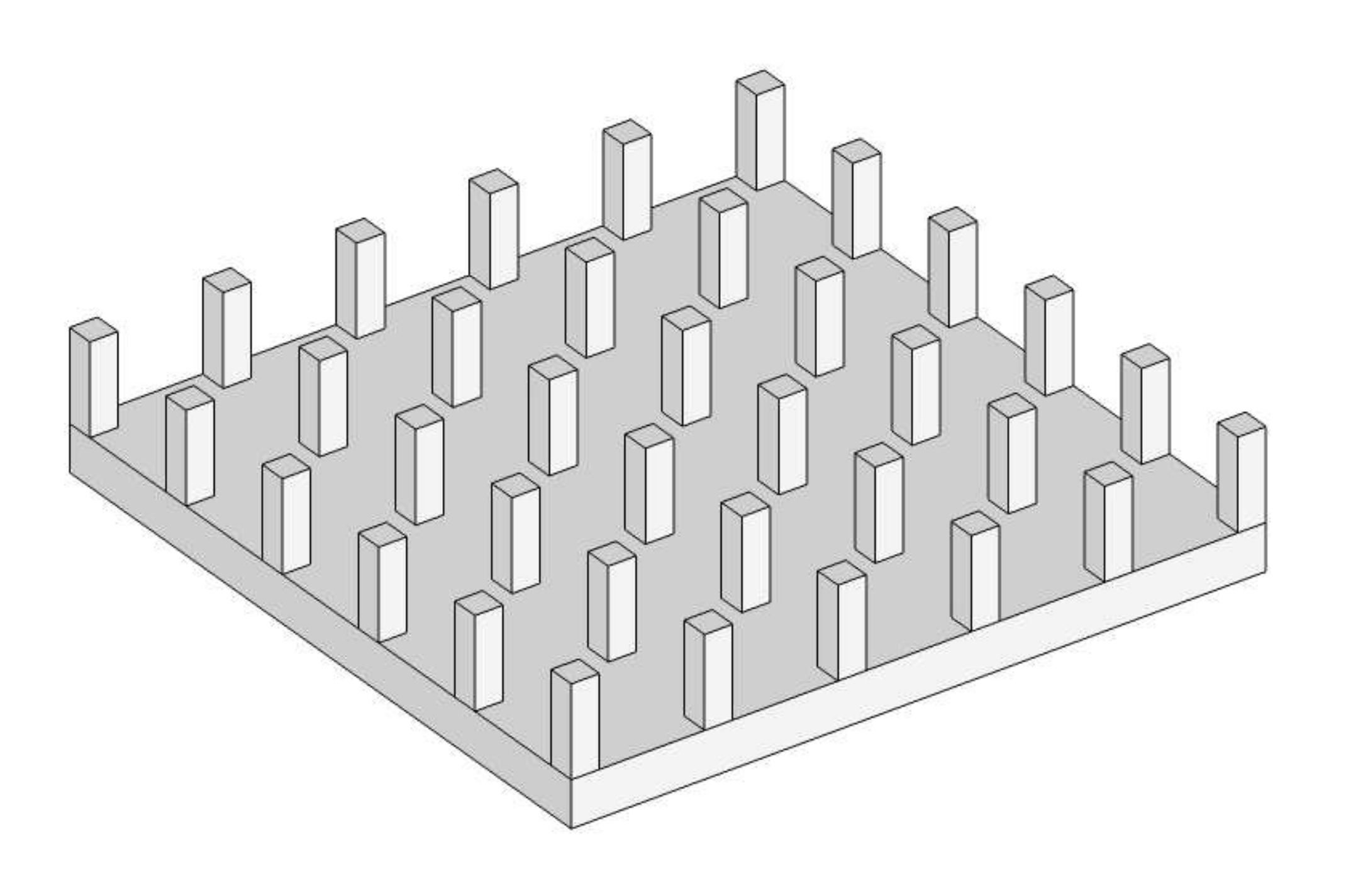}
    \caption{}
    \label{3D_sup1}
  \end{subfigure}\\[1ex]
  \begin{subfigure}{\linewidth}
    \centering
    \includegraphics[width=.95\linewidth]{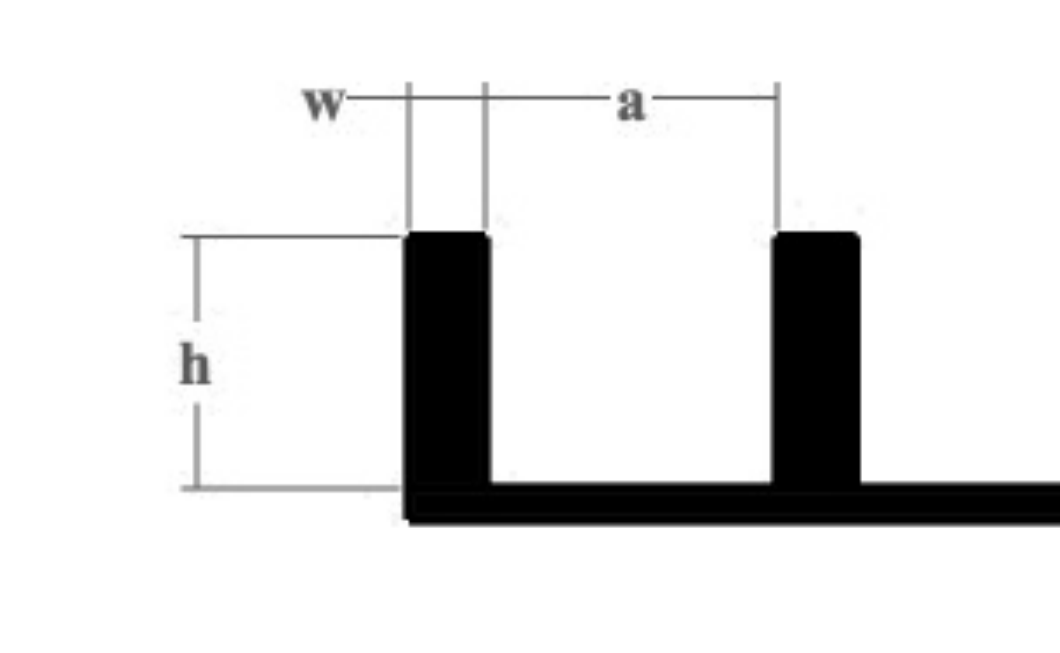}
    \caption{}
    \label{2D_sup1}
  \end{subfigure}
\end{minipage}%
\begin{minipage}{.33\textwidth}
  \begin{subfigure}{\linewidth}
    \centering
    \includegraphics[width=.95\linewidth]{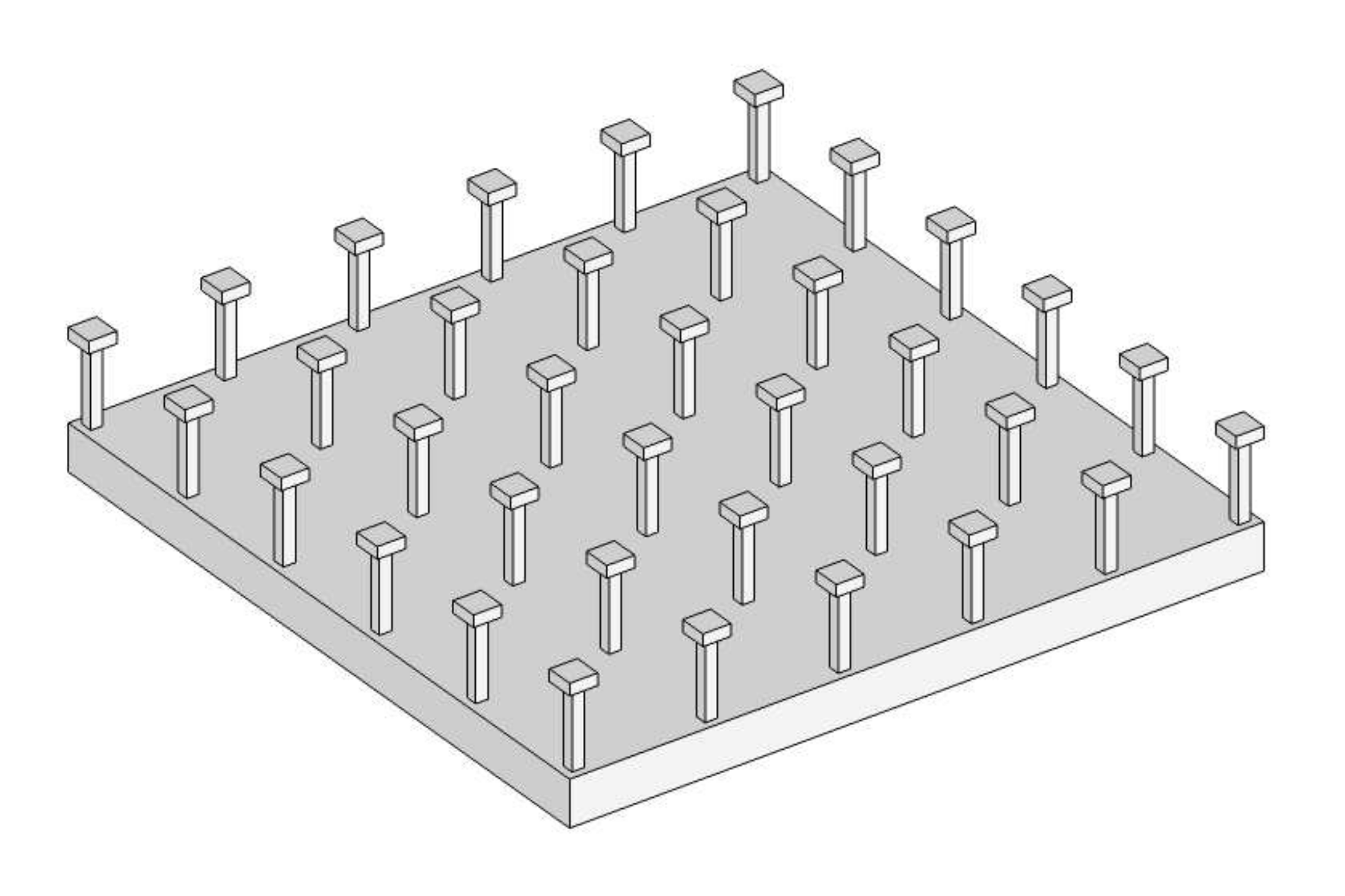}
    \caption{}
    \label{3D_sup2}
  \end{subfigure}\\[1ex]
  \begin{subfigure}{\linewidth}
    \centering
    \includegraphics[width=.95\linewidth]{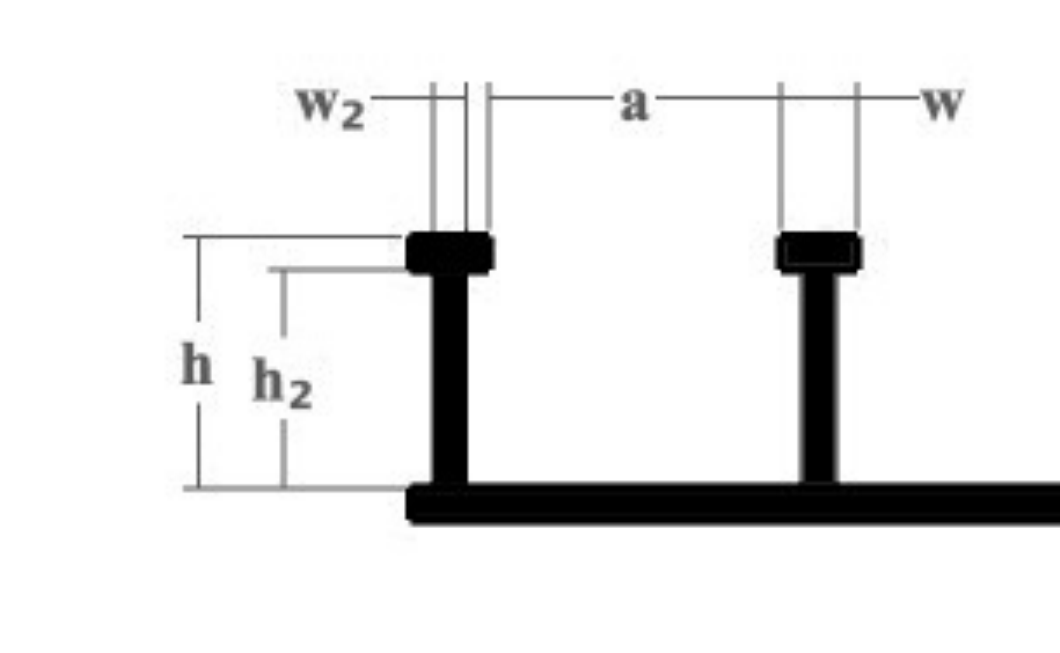}
    \caption{}
    \label{2D_sup2}
  \end{subfigure}
\end{minipage}%
\begin{minipage}{.33\textwidth}
  \begin{subfigure}{\linewidth}
    \centering
    \includegraphics[width=.95\linewidth]{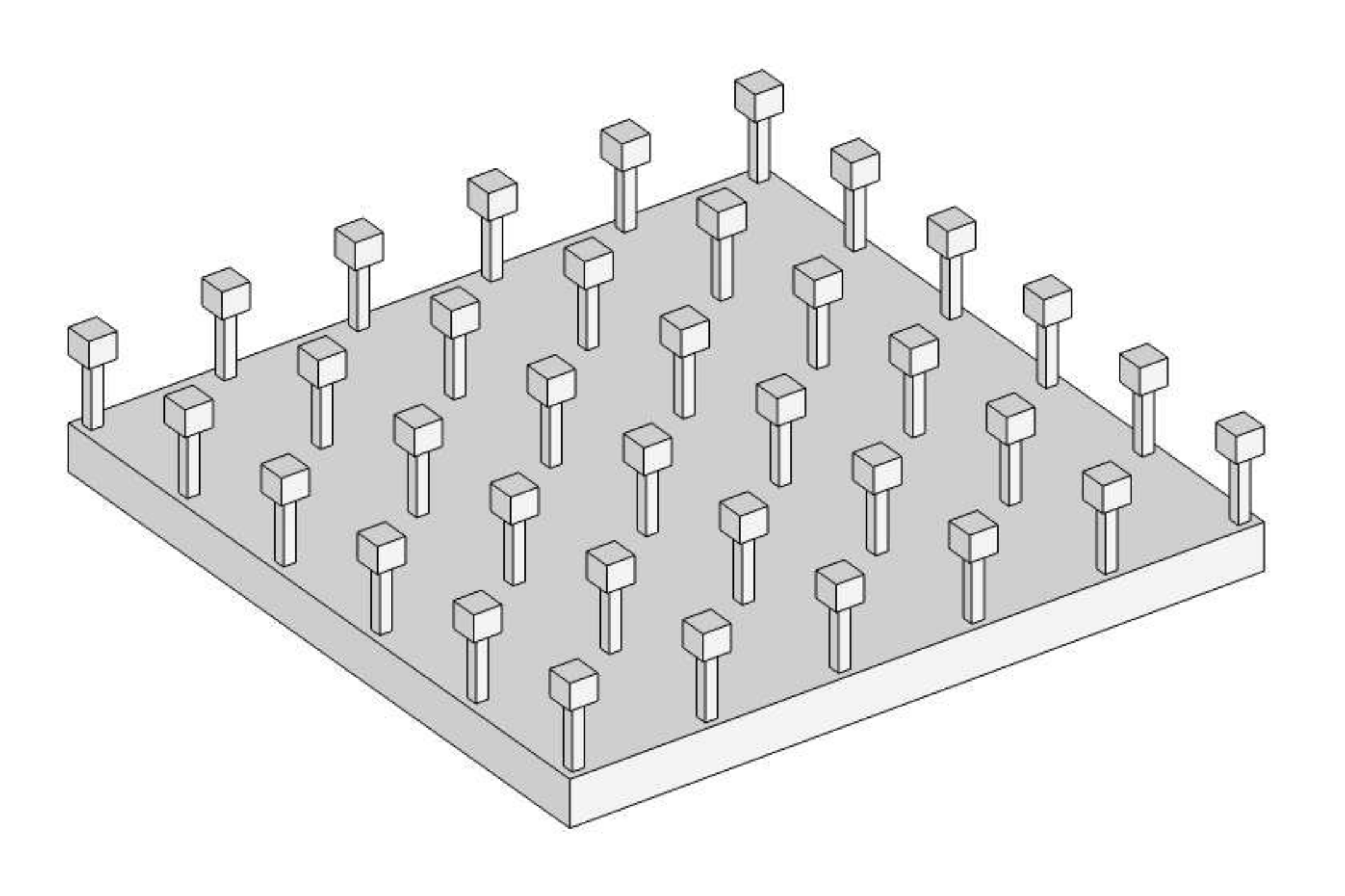}
    \caption{}
    \label{3D_sup3}
  \end{subfigure}\\[1ex]
  \begin{subfigure}{\linewidth}
    \centering
    \includegraphics[width=.95\linewidth]{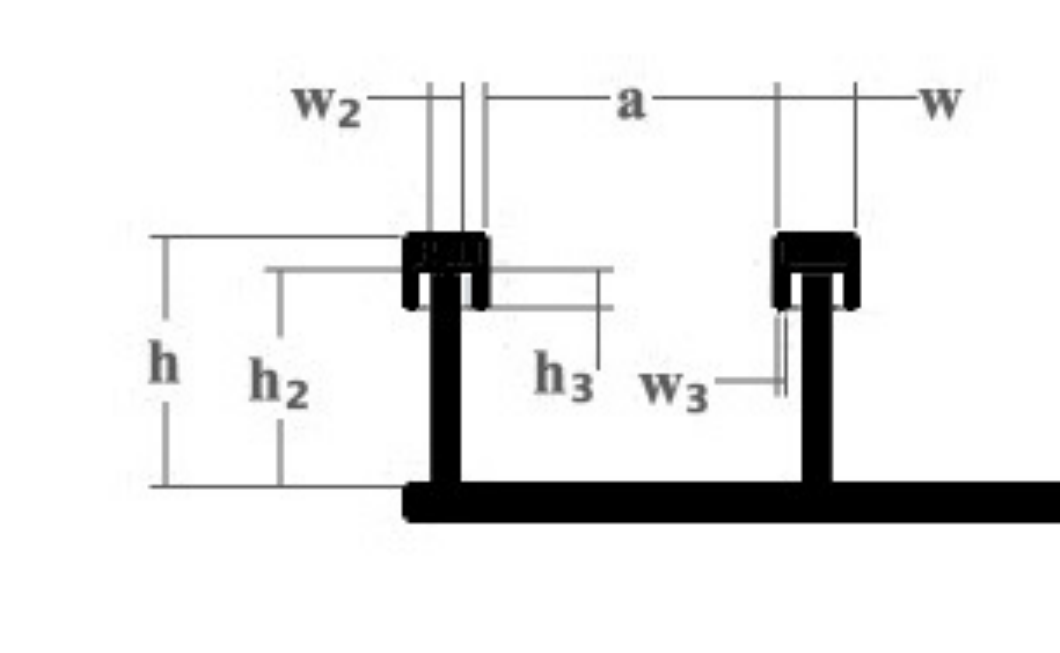}
    \caption{}
    \label{2D_sup3}
  \end{subfigure}
\end{minipage}
\caption{ Surfaces analyzed in this work. {\bf (a)} Schema 3D of the {\it pillared surface}, also called {\it surface of type 1}.  {\bf (b)} 2D section of the pillared surface and the definitions of its geometric parameters: pillars height $\hh$, distance between pillars represented by $\aaa$ and pillar width given by $\ww$. {\bf (c)} Schema 3D of the  {\it simple reentrant  surface}, also referred as  {\it surface of type 2}. {\bf (d)} Definition of its geometric parameters: the basis of the pillars are decreased, possessing  width  $\ww_2 \in (0,\ww)$ and height $\hh_2 \in (0,\hh)$ and creating an horizontal overhang  as shown in the figure.   {\bf (e)} Schema 3D of the  {\it double reentrant  surface} or {\it surface of type 3}. {\bf (f)} Definition of the geometric parameters:  this surface is built by adding a vertical overhang with length $\hh_3 \in (0, \hh_2)$ and thickness $\ww_3 \in (0, (\ww-\ww_2)/2)$, generating a double reentrance. }
\label{typesSurface}
\end{figure}

\begin{figure}
\centering
\includegraphics[width = 0.33\columnwidth]{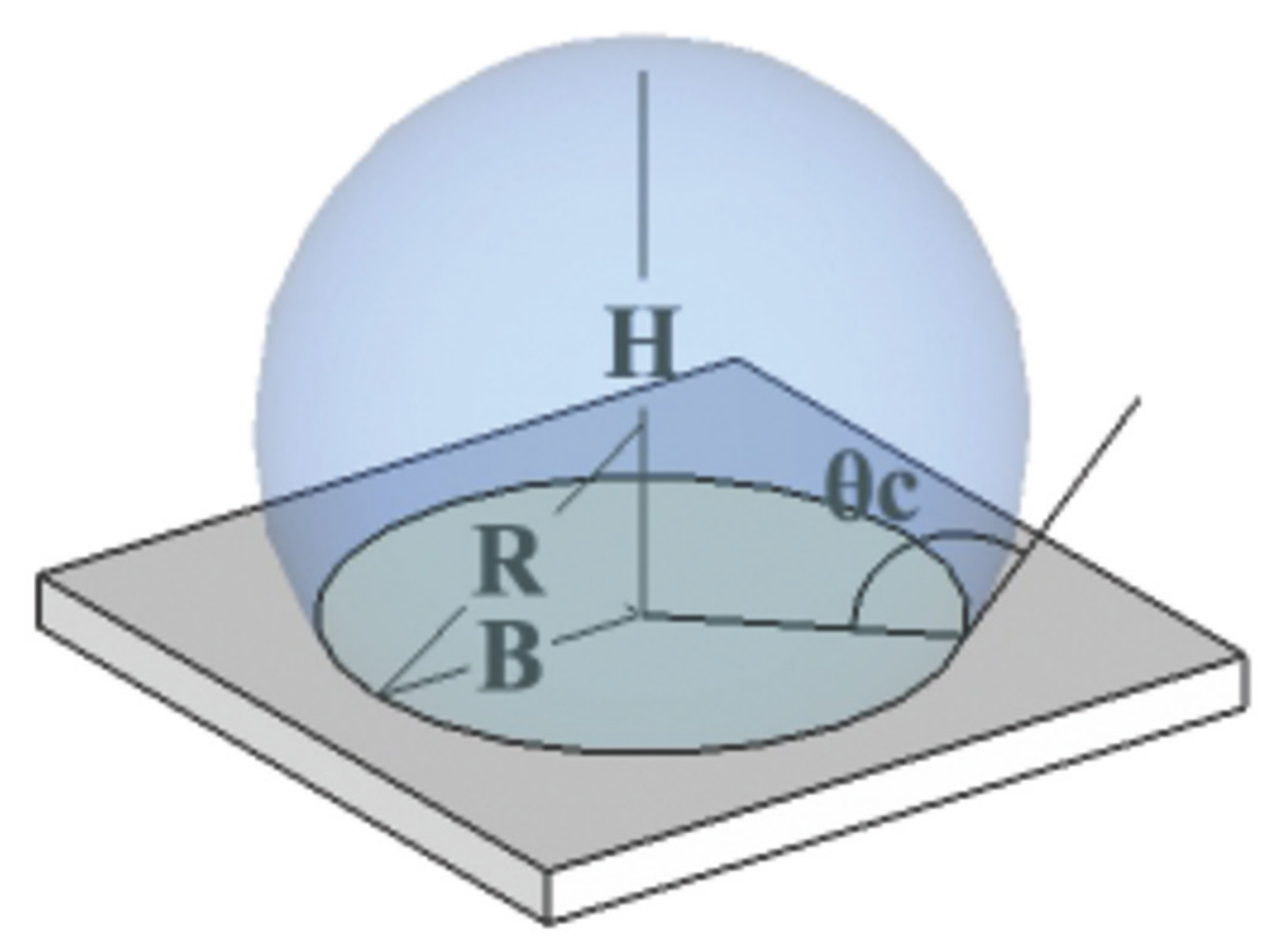}
\caption{Geometric parameters of the droplet. We consider that a stable droplet assumes the shape of a spherical cap with radius $R$, base radius $B$, height $H$ and contact angle $\theta_C$.}
\label{geometryDroplet}
\end{figure}

In this section we develop a model that takes into account the energy cost of creating interfaces between different phases when a droplet of a given volume $V_0$ is placed on a  surface of three types, as  schematized  in Fig.(\ref{typesSurface}). The model and the method used to minimize the global energy were developed  in a previous work \cite{Fernandes2015} to study the case where a droplet is placed on a surface of type 1, Fig.(\ref{3D_sup1}). Here we extend the method for the reentrant and double-reentrant surfaces, as the ones shown in Fig.(\ref{3D_sup2}) and Fig.(\ref{3D_sup3}).

We consider a three dimensional spherical droplet with geometric parameters as defined in Fig.(\ref{geometryDroplet}). The droplet is supposed to be in one of the two possible states, the Cassie-Baxter (CB)  or the Wenzel (W) state.
We emphasize that in this work we consider one particular case of the CB state, which is the Fakir configuration with no liquid penetrating the surface. The W state considered here is the homogeneous one, where the liquid fully penetrates the grooves.
The total energy of each state is given by the sum of all energies involved in creating interfaces between every pair formed from  liquid, solid, and  gas after the droplet is placed on a surface,  $E_{\bf int}^{\rm s}$. This energy is subtracted from the energy of the surface without the droplet, $E_{\bf surf}$, and  the relevant quantity to define how much energy a given state $s$ ($s$=W or $s$=CB) costs is the difference $ \Delta E^{\rm s}=E_{\bf int}^{\rm s}-E_{\bf surf}$. For the droplet sizes considered in this work, the gravitational energy of the droplet is of order of $10^{-4}$ times the interfacial energy and it can be safely neglected.

In the CB state the droplet only touches the surface on the top of the pillars, which size is given by $\ww ^2$ for all the tree types of surfaces, as indicated in Fig.(\ref{typesSurface}). Because there is no liquid in the internal part of the surface, the energy of the droplet in the CB state is the same for the three types of surface. 
Using Young's relation, $\sigma_{SG} - \sigma_{SL} = \sGL \cos(\tty)$, we can write the energy of the CB state as: 
\begin{eqnarray}
\Delta E^{\bf {CB}} &=&  \sGL \left[ N^{\rm CB} ~( \dd^2 - \ww^2 (1 + \cos\theta_Y) ) +  S ^{\rm CB} \right], \label{en_CB}
\end{eqnarray}
where $\dd= \ww + a$  and $\sGL$ is the liquid-gas interfacial tension.
The total number of pillars underneath the droplet is
 $ N^{\rm s} = \frac{\pi}{4}(2 B ^{\rm s}/ \dd)^2$, where 
$B^{\rm s} = R^{\rm s}\sin(\ttc^{\rm s})$  is the base radius. The surface area of the spherical cap in contact with air is given by 
$S^{\rm s}=2\pi  {R^{\rm s}}^2 [1-\cos (\ttc^{\rm s})]$. 

 On the other side, in the W state the droplet is in contact with the internal  part of the surface and  therefore the energy terms will be different for each kind of surface:

\begin{eqnarray}
\Delta E_{\bf (1)}^{\rm {W}} &=&  \sGL [ S^{\rm W}_{\bf (1)} - {\rm N}^{\rm W}_{\bf (1)}  (  \underbrace{\dd ^2 + 4 \ww \hh}_{T_1} )\cos \tty ], \label{en_W_s1} \\
\Delta E_{\bf (2)}^{\rm {W}} &=&  \sGL [ S^{\rm W}_{\bf (2)} - {\rm N}^{\rm W}_{\bf (2)}  (\dd ^2 + 4 \ww \hh + \underbrace{2(\ww^2-\ww_2^2) - 4 \hh_2(\ww - \ww_2)}_{T_2} ) \cos \tty ] , \label{en_W_s2}\\
\Delta E_{\bf (3)}^{\rm {W}} &=&  \sGL [ S^{\rm W}_{\bf (3)} - {\rm N}^{\rm W}_{\bf (3)}  (\dd ^2 + 4 \ww \hh + 2(\ww^2-\ww_2^2) - 4\hh_2 (\ww - \ww_2) +\underbrace{4 \hh_3 (2\ww - \ww_3)}_{T_3} ) \cos \tty ], ~~~~\label{en_W_s3}
\end{eqnarray}
where the subscript $1, 2, 3$ indicate the indexes of the three types of surfaces.  We remind that all the geometric parameters are defined in Fig.(\ref{typesSurface}) and Fig.(\ref{geometryDroplet}).

To define which wetting state is stable, W or CB, we find the minimum energy state for a given geometry and type of liquid. To do so,  we do not use the absolute values of the energies, but only their difference. Because  surface tension of the liquid  $\sGL$ multiplies all the equations above, this term does not influence the thermodynamic stable state. Therefore we will assume that $\sGL=1$ and the only information about the type of the liquid in the model is contained in  $\tty$.

In what follows we discuss some analytical limits of these equations that guide us to compare the energies of the droplet in the different surfaces. At the end of this section we explain the minimization procedure used to define the wetting stable state and how to obtain the wetting diagram for a droplet placed on these three surfaces.

\subsection*{Theoretical considerations about the model ~~}

In this section we consider a limit case where the radius of the droplet is large compared to the typical scale of roughness. In this limit the volume of the liquid inside the roughness groves is negligible compared to the volume of the cap. Then $ N^{\rm s}$  and $S^{\rm s}$ are the same for all the surfaces and the expressions of energies can be rewritten as:
\begin{eqnarray*}
\Delta E^{\bf {CB}} &=&  S + N ~( \dd^2 - \ww^2 (1 + \cos\theta_Y) ), \label{en_CBb} \\
\Delta E_{\bf (1)}^{\rm {W}} &=& S - {\rm N} ~ {T_1} \cos \tty , \label{en_W_s1b} \\
\Delta E_{\bf (2)}^{\rm {W}} &=& \Delta E_{\bf (1)}^{\rm {W}} - {\rm N}~{T_2} \cos \tty , \label{en_W_s2b} \\
\Delta E_{\bf (3)}^{\rm {W}} &=& \Delta E_{\bf (2)}^{\rm {W}} - {\rm N}~{T_3} \cos \tty , \label{en_W_s3b}
\end{eqnarray*}
with $T_1, T_2, T_3$ defined in the Eq.(\ref{en_W_s1}) - Eq.(\ref{en_W_s3}). Note that $T_2$ can be zero, which happens when $ \hh_2=\hh_2^{\ast}=\frac{(\ww + \ww_2)}{2}$, positive or  negative.  $T_3$ is always positive but the value of $\hh_3$ does determine the relation between the energies of the surfaces.  The parameter  $\hh_3^*=[\hh_2-\hh_2^*]\frac{(\ww-\ww_2)}{(2\ww-\ww_3)}$ is determinant in defining these relations (see below).

The first question is about the possibility of CB being the lowest energy state.
For the case where $\tty>90^{\circ}$ it is possible mathematically the relation $\Delta E^{\bf {CB}} < \Delta E_{\bf (i)}^{\rm {W}}$ for the three types of surfaces, $\text{i}=1,2,3$. 
It implies that in this situation the thermodynamic stable state of the droplet can be the CB for all the three types of surfaces depending on its  geometric parameters.
However, for the case $\tty<90^{\circ}$, there is no set of geometric parameters for any of the type of surfaces considered in this work for which one could build a CB as the stable state. In terms of energy, it means that  $\Delta E^{\bf {CB}} > \Delta E_{\bf (i)}^{\rm {W}}$ always.

Another question to address is which interval of geometric parameters increases the energy of the W state  when changing the type of surface.
Note that even in the cases where the CB state is not reachable, the fact that the energy of the W state increases implies that the contact angle of the droplet has a chance to increase as well. 
In other words, to find the conditions for which $ \Delta E_{\bf (i)}^{\rm {W}}$ increases is related to the possibility of enhancing   $\ttc$ of the droplet. The enhancement of $\ttc$ is associated with the repellency of the surface: higher is $\ttc$, more repellent is the surface. This argument does not take into account the energy barrier which is known to be important in this phenomenology \cite{TutejaSicence2007,joly2009,Escobedo2012a} and will be discussed in a next section.

Table(\ref{tabComparaE}) shows a comparison between the W energies of the three types of surfaces,  indicating which are the  interval of geometric parameters that increase the energy of the W state.
We then take into account the inequalities shown in the Table(\ref{tabComparaE}) and build the Table(\ref{tabComparaE2}) with all the theoretical possible relations between the contact angle $\ttc$ of the droplet placed on the surfaces and the geometric conditions for all of these situations.
$\theta_i$ means the contact angle of the thermodynamically stable state of the droplet on the surface of type $i=1,2,3$.  Below Table(\ref{tabComparaE2}) it is  shown a schema of the geometric configurations that represents each condition for the case $\tty> 90^{\circ}$. 

\begin{table*}
\begin{tabular}{|l||c|c|c||} 
  \hline
    & ${\bf \Delta E_{\bf (2)}^{\rm {W}}>\Delta E_{\bf (1)}^{\rm {W}}}$ &  ${\bf \Delta E_{\bf (3)}^{\rm {W}}>\Delta E_{\bf (2)}^{\rm {W}}}$ &  ${\bf \Delta E_{\bf (3)}^{\rm {W}}>\Delta E_{\bf (1)}^{\rm {W}}}$  \\ \hline \hline
$\tty> 90^{\circ}$ &      $\hh_2 < \hh_2^*$ &    always    &  \makecell{{\bf (a)}~ $\hh_2 < \hh_2^{\ast}$ {\it or}  \\ {\bf (b)}~  $\hh_2 > \hh_2^{\ast}$ ~and~ $\hh_3>\hh_3^*$ } \\ \hline
 $\tty< 90^{\circ}$    &     $\hh_2 > \hh_2^{\ast}$    &    never  &   $\hh_2 > \hh_2^{\ast}$ ~and~ $\hh_3<\hh_3^*$ \\ \hline
   \hline
\end{tabular}	
\caption{It shows the interval of geometric parameters for which the energies of the droplet placed on different surfaces  present global energies for the W state as given by the relation shown on the top of each column. We remind the definition of $\hh_2^{\ast}=\frac{(\ww + \ww_2)}{2}$ and $\hh_3^*=[\hh_2-\hh_2^*]\frac{(\ww-\ww_2)}{(2\ww-\ww_3)}$. Note that for the case ${\bf \Delta E_{\bf (3)}^{\rm {W}}>\Delta E_{\bf (1)}^{\rm {W}}}$ and $\tty> 90^{\circ}$ there are two different conditions, denoted by {\bf (a), (b)}.}
\label{tabComparaE}
\end{table*}

\newcolumntype{C}{>{\centering\arraybackslash}p{9.5em}}
\begin{table}
\centering
\begin{tabular}{||C|C||C||} 
\hline
\multicolumn{2}{||c||}{\makecell{Relations between the $\ttc$ of the surfaces}}&\multicolumn{1}{c||}{\makecell{Geometric condition}} \\
 $ \centering \tty<90^{\circ} $  & $\tty>90^{\circ}$ &  \\ \hline \hline
  $\theta_1 > \theta_2> \theta_3$  & {\bf (a)}  $\theta_3 > \theta_2> \theta_1$ & $\hh_2 < \hh_2^{\ast}$  \\ \hline
  $\theta_2> \theta_3> \theta_1$ & {\bf (b)}  $\theta_1> \theta_3> \theta_2$   & $\hh_2 > \hh_2^{\ast}$ ~and~ $\hh_3<\hh_3^*$  \\ \hline
  $\theta_2> \theta_1> \theta_3$ & {\bf (c)}  $\theta_3> \theta_1> \theta_2$   & $\hh_2 > \hh_2^{\ast}$ ~and~ $\hh_3>\hh_3^*$ \\ \hline
  $\theta_2 >  \theta_1 =\theta_3$ & {\bf (d)}  $\theta_3= \theta_1> \theta_2$   & $\hh_2 > \hh_2^{\ast}$ ~and~ $\hh_3=\hh_3^*$ \\\hline
  $\theta_2= \theta_1 >  \theta_3$  & {\bf (e)}  $\theta_3> \theta_2= \theta_1$   & $\hh_2 = \hh_2^{\ast}$ \\ \hline
\hline 
\end{tabular}	
\includegraphics[width=1\textwidth]{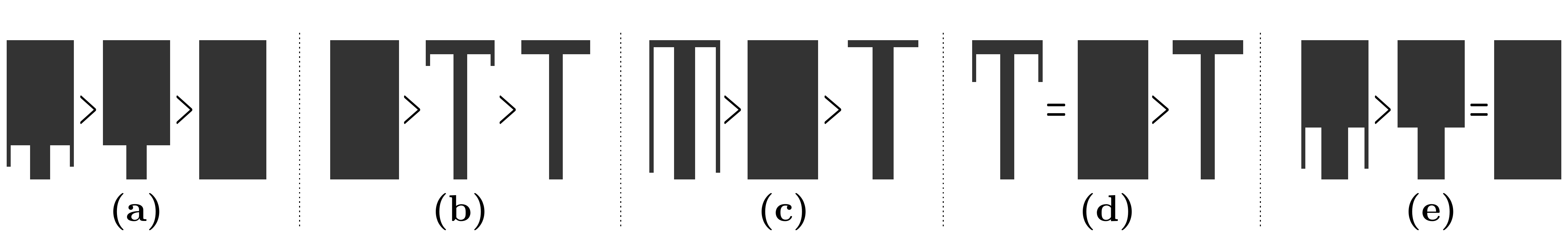}
\caption{The table summarizes all possible mathematical relations between the $\ttc$ for the three surfaces and its respective geometric conditions divided in the two cases $\tty<90^{\circ}$ and $\tty>90^{\circ}$. Below the table there is a schema of the surfaces for each of the five geometric conditions. The symbols refer to the relations between $\ttc$ of the droplet on the different types of surfaces.}
\label{tabComparaE2}
\end{table}

The table and the figures indicate non-trivial relations between the geometric parameters of the surfaces and the result in terms of the contact angle of the droplet. The relation denoted by "a",  where  $\theta_3 > \theta_2> \theta_1$, corresponds to a geometrical situation where $\hh_2 < \hh_2^{\ast}$ , with $\hh_2^{\ast}=\frac{(\ww + \ww_2)}{2}$.   Besides the fact that $\hh_2$ depends on the  widths of the reentrances and not on their heights, the result is such that the height of the simple reentrance is small.  There is no condition on the overhang of the second reentrance to create this situation.   The situation "b,c,d" happens when $\hh_2 > \hh_2^*$,  but depending on the value of the overhang $\hh_3$ there are 3 possibilities as shown in the schema.  Situation "e" happens when the term $T_2=0$  in the Eq.(\ref{en_W_s2}) is equals to zero and mathematically there is no effect of the first reentrance.

It is important to realize that the analysis of the equations developed in this section allow us to understand the range of parameters for which the energy of one state can overcome the energy of the other state or can enhance the contact angle of the droplet. These analysis cannot, however, predict which is the value of the apparent contact angle $\ttc$ of the thermodynamically stable state of the droplet on each type of surface.
To do so, one needs to implement the minimization procedure explained in the next section.

It is worth noting that in the experiments where the droplet evaporates, eventually the volume of the droplet becomes small compared to the typical scale of roughness and  a transition from CB to W is observed~\cite{Tsai2010, Xu2012, Fernandes2015, Tsai2017}. Is these cases,  the volume below the grooves can compete with the term of the cap and some considerations made above can fail \cite{gao2007wenzel, Marmur_CWright, gao2009attempt}. 

\subsection*{ Energy minimization ~~}

To decide which wetting state (W or CB) is favorable from the thermodynamic point of view, we minimize the equations of global energy derived above and compare the minimal energy for each state. This minimization procedure  was discussed  in the reference \cite{Fernandes2015} for the pillared surface. Here we recall the idea for a surface of type 1 and apply the method for the types of surface 2 and 3. In the Supporting Information (SI) we show a flowchart, Fig.(S1), of the method and explain how to extend it for surfaces of type 2 and 3. 

Consider a surface of type 1. We fix all its geometric parameters $\hh$, $\aaa$ and $\ww$  and its chemical properties (in practice, we only need to chose $\tty$) and ask the following question: if a droplet of a fixed volume $V_0$ is placed on this surface, which would be its final wetting state, W or CB?
If the geometry and $\tty$ are fixed,  the energies expressed in Eq.(\ref{en_CB}) and Eq.(\ref{en_W_s1}) only depend on the droplet radius $R^{\rm s}$ and on the contact angle  $\ttc^{\rm s}$.  

To find the minimum of CB and of the W state, we do the following: 
(i) we compute the radius $R^{\rm s}$ by solving the cubic equation for a fixed volume $V_0$ (the equations for the volume of each surface are shown in the SI). (ii) Then, we vary the contact angle $\ttc^{\rm s} \in (0, \pi]$ and for each contact angle, we compute the energy difference  $\Delta E^{\rm s }$ associated with these parameters using Eqs.~(\ref{en_CB}) and (\ref{en_W_s1}). (iii) We compare all the energies found for  $\Delta E^{\bf {CB}}$ and store the minimum one, called  $\Delta E^{\bf {CB}}_{\bf min}$. 
There is one detail in this step:  to select $\Delta E^{\rm {CB} }_{\bf min}$, we also impose the  constraint that the contact line of the droplet has to be pinned to the pillars~\cite{Shahraz2012}. This implies that the base radius $B^{\rm {CB}}$ and $\theta^{\rm {CB}}$ does not have a continuous value as a function of volume.
We do the same for the W state and define  $\Delta E^{\bf {W}}_{\bf min}$. (iv)  The thermodynamically stable state is the one with the lowest $\Delta E^{s}$. In other words,  if $\Delta E^{\bf {W}}_{\bf min} < \Delta E^{\bf {CB}}_{\bf min}$, the W is the stable state.

Once the state with the lowest energy  is defined, all geometric parameters of the droplet (contact angle $\ttc$, radius $R$, base radius $B$, spherical cap height $H$) in this state are determined.  
 This procedure can be applied for any set of geometric parameters $(\hh, \aaa, \ww)$ and value of  $\tty$ to build the wetting diagram for the pillared surface.

\section*{Theoretical Results and Discussion}

In the previous section we discussed the theoretical possibilities for the energies of the droplet placed on each type of surface and we observed that, depending on the geometric parameters, there are five possible relations between the $\ttc$ on different surfaces, summarized in Table(\ref{tabComparaE2}). These relations guide us to look for the enhancement of the $\ttc$, but to know {\it by how much} is the $\ttc$ enhanced we need to apply the minimization procedure explained before.
Our goal in this section is to explore the wetting diagrams of all types of surfaces, focusing in quantifying when the  difference in $\ttc$ is maximized for the three types of  surfaces.

\subsection*{Pillared {\it vs} simple reentrant surfaces}

\begin{figure}[H]
\centering
\includegraphics[width = 1.0\columnwidth]{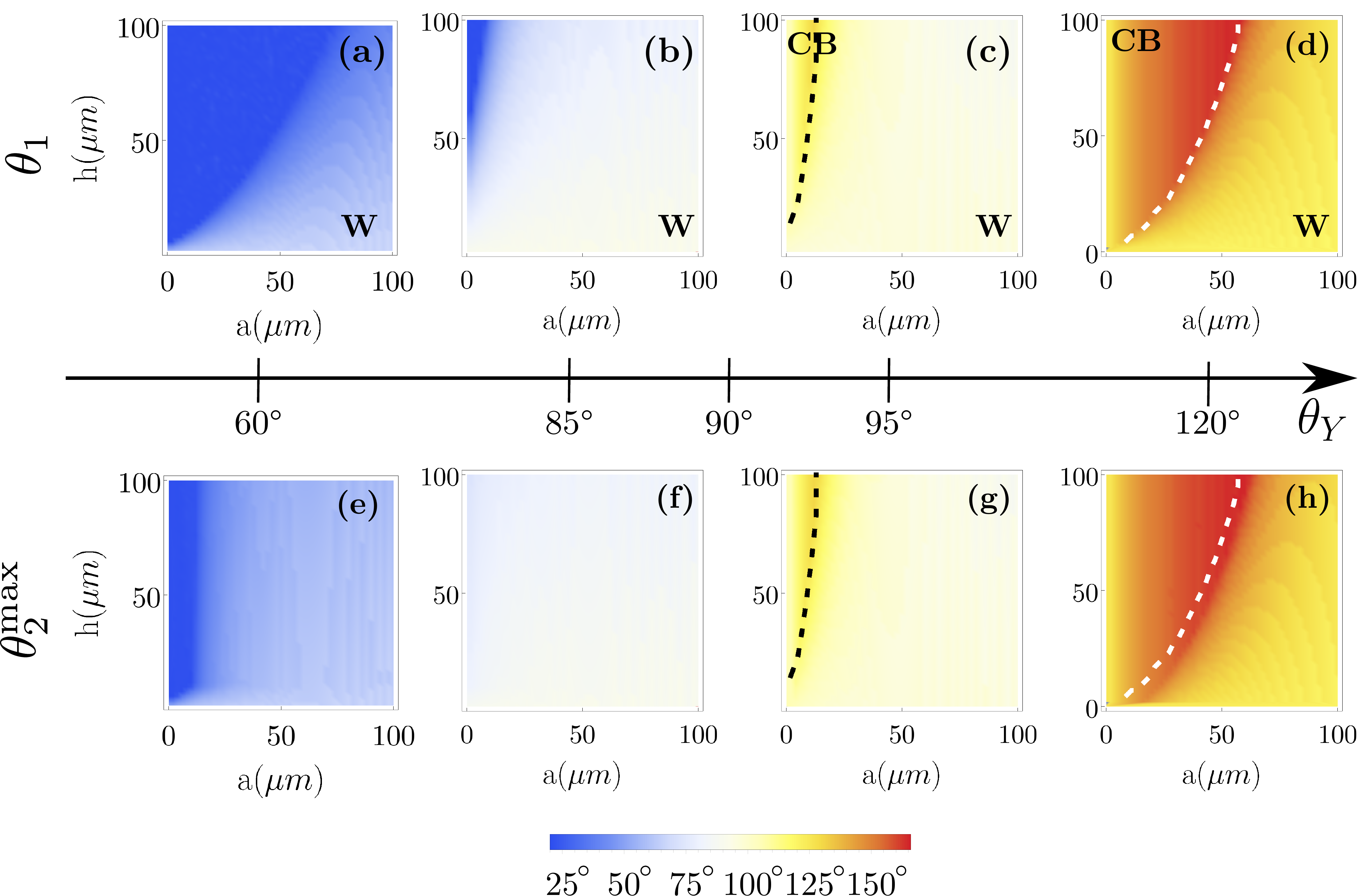}
\caption{{\bf(a)-(d)}: Wetting diagrams for the surface of type 1. The quantity shown is the contact angle $\theta_1$ for a droplet radius   $R_0= 1000 \mu $m as a function of two geometric parameters of the surface: the height of the pillars $\hh$ and the distance $\aaa$ between them. The dotted line, when it appears, shows the predicted thermodynamic transition between the Cassie-Baxter and Wenzel states, being that the Wenzel state corresponds to the region below the line. In the case where $\tty<90^{\circ}$, there is no thermodynamic transition. {\bf(e)-(h)}: diagram of $\theta^{\mm}_2$ as defined in the text. From the left to the right the $\theta_Y$ is increased, ranging from a wetting to non-wetting case. $\ww=20\mu m$ for all diagrams.}
\label{compS1S2}
\end{figure}

The wetting diagrams for the surface of type 1 are shown in Fig.(\ref{compS1S2})-a,b,c,d. These are diagrams of the contact angle of the most  stable wetting state when the droplet placed on a surface of type 1, named $\theta_1$,  as a function of the pillar height $\hh$ and pillar distance $\aaa$  for several  $\tty$. To build these diagrams, we fix the $R=1000\mu$m (corresponding to a volume $V_0=4,2\mu $l), the value of $\theta_Y$ and then, for each set of parameters $(\hh, \aaa, \ww)$, the equations Eq.(\ref{en_CB}) and Eq.(\ref{en_W_s1}) are minimized. 

When $\tty>90^\circ$, the CB state is the thermodynamic stable state for small values of $\aaa$ and high values of $\hh$ and there is a transition to the W state when $\aaa$ increases and $\hh$ decreases, as the dashed line indicates in Fig.(\ref{compS1S2})-c,d \cite{Fernandes2015}.  
When $\theta_Y$ decreases, the CB region also decreases gradually and disappears for $\theta_Y=90^{\circ}$. Below this value, there is no  transition: the only stable thermodynamic state is the Wenzel state.

What are the geometries that maximize the enhancement of $\ttc$ on the surface of type 2 compared to this value on the surface of type 1?
To answer to this question, one needs to span systematically all the geometric parameters that define the surface of type 2. To do so, we developed the following procedure. (i) We fix a set of parameters $(\hh, \aaa, \ww)$ of the surface of type 1 and vary the parameters of the surface of type 2 taking into account all the possibilities $\ww_2 \in (0, \ww)$  and $\hh_2 \in (0, \hh)$.  (ii) For each set of parameters $(\hh, \aaa, \ww, \ww_2, \hh_2)$ of the surface of type 2, we minimize  Eq.(\ref{en_CB}) and Eq.(\ref{en_W_s2}) and find the contact angle
$\theta_2$ that minimizes the global energy of the droplet on this surface 2, as explained in the SI. (iii) After spanning all possible geometries of the surface 2, we search for 
$\theta_2^{\mm}$ which is defined as the angle that maximizes the difference between $\theta_2$ and $\theta_1$. 
We refer to the surface that produces $\theta_2^{\mm}$ as an {\it optimal surface} for the set of parameters $(\hh, \aaa, \ww)$ and the geometric parameters responsible for that as $\ww_{2}^{\rm opt}$ and $\hh_{2}^{\rm opt}$.

Fig.(\ref{compS1S2})-e,f,g,h show $\theta_2^{\mm}$ for different values of $\theta_Y$. In the case where $\tty>90^{\circ}$, comparing diagrams (c) and (g) for $\tty=95^{\circ}$  or diagrams (d) and (h) for $\tty=120^{\circ}$, we observe typically no difference between $\theta_1$ and $\theta_2^{\mm}$.
In the case where $\tty<90^{\circ}$, comparing for example the diagrams (a) and (e) for $\tty=60^{\circ}$ we also observe that for some region the contact angle increases from the surface 1 to the surface 2.

Surface roughness $r$ for the surface of type 1 is defined as $r_1 =1+ ( 4 \hh \ww )/\dd^2$  and for the surface of type 2 the definition is given by  $r_2 - r_1= (2(\ww^2-\ww_2^2) -4\hh_2(\ww-\ww_2))/\dd^2$, with $\dd=\aaa+\ww$. 
In the case of $\tty>90$, typically  $\ww_{2}^{\rm opt}\rightarrow 0$ and $\hh_2^{\rm opt} \rightarrow 0$ (both values are finite because we impose minimal values for these parameters), which results in $r_2^{\rm opt}-r_1= 2\ww^2/\dd^2$, where $r_2^{\rm opt}$ is the roughness of the optimal surface. We can analyze the variation of $r_2^{\rm opt}$ in respect to $r_1$: for small values of $\aaa$, $r_2^{\rm opt}- r_1 \approx 2$, while for big values of $\aaa$, $r_2^{\rm opt} \approx r_1$.
The fact that $r_2^{\rm opt}$ is similar to $r_1$  agrees with  $\theta_2^{\mm}\approx \theta_1$.
In the case  $\tty<90^{\circ}$, typical values are  $\ww_{2,\rm opt}\rightarrow 0$ and $\hh_{2}^{\rm opt} \rightarrow \hh_2$, implying that $r_2^{\rm opt} -r_1 = 2\ww(\ww -2\hh_2^{\rm opt})/\dd^2$. 
For most of the points of the diagram, $r_2^{\rm opt}< r_1$ by a factor that depends on $(\ww -2\hh_2^{\rm opt})/\dd^2$. 
For example in the case of the Fig.(\ref{compS1S2}) and $\tty=85^{\circ}$, the value of  $r_2^{\rm opt} -r_1$ vary from 0 to -12, which  agrees with our measure $\theta_2^{\mm} > \theta_1$. 

\subsection*{Pillared,  simple and double reentrant surfaces}

In this section we consider the surfaces with double reentrance,  Fig.(\ref{3D_sup3}).
From Table(\ref{tabComparaE2}) we note that the global minimum energy contact angle of a droplet placed on a surface of type 3, $\theta_3$, is the highest in most of the geometric situations for the cases where $\tty>90^{\circ}$. However, $\theta_3$ is not the highest contact angle in any of the geometric parameters for the case  $\tty < 90^{\circ}$.
For this reason we only analyze the situation where $\tty>90^{\circ}$, setting $\tty=120^{\circ}$.

\begin{figure}
\includegraphics[width = 1\columnwidth]{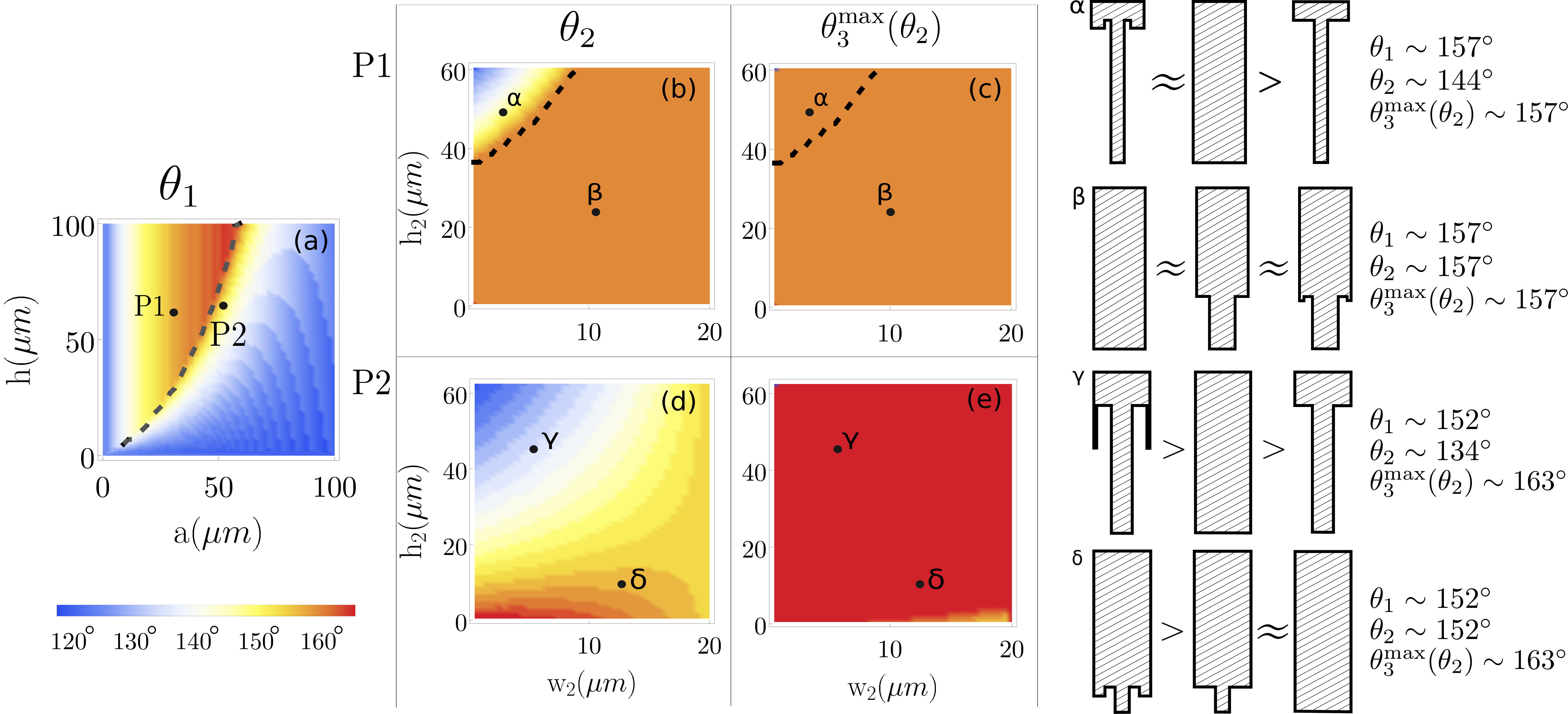}
\caption{{\bf (a)} $\theta_1$ as a function of the height of the pillars $\hh$ and the distance $a$ between them. {\bf (b, d)} Diagrams $\theta_2$ as a function of the height $\hh_2$ and the width $\ww_2$. {\bf (c,e)} Diagrams of $\theta^{\mm}_3(\theta_2)$  as a function of the height $\hh_2$ and the width $\ww_2$ of the simple reentrance. 
For the diagrams (b,c) the values of $\hh$ and  $a$ are given by the point $P_1$ of the diagram (a) and for the diagrams (d,e) the values of $\hh$ and  $\aaa$ are given by the point $P_2$ of the diagram (a).
The figures on the right are schema of the surfaces with the correct proportion between the geometric parameters. Each set of 3 surfaces are referent to the points $\alpha, \beta, \gamma, \delta$ marked on the diagrams (b,c,d,e). On the right of each set of figures is indicated the values of $\theta_1$, $\theta_2$ and $\theta^{\mm}_3$ for each point. $\ww=20\,\mu m$,  $R_0=1000\,\mu$m  and $\tty=120^{\circ}$ for all diagrams.}
\label{compS1S2S3}
\end{figure}

Fig.(\ref{compS1S2S3})-a shows the diagram of the $\theta_1$, which was shown in Fig.(\ref{compS1S2}) but it is repeated here to indicate the points $P_1$ and $P_2$  that are analyzed in detail. Note that $P_1$ is in the CB state, $P_2$ is in the W state and both are close to the transition line. We will refer to the set of geometric parameters that defines these points $P_\text{i}$ as ($\aaa^{P_\text{i}}, \hh^{P_\text{i}}, \ww^{P_\text{i}})$. At the end of this section we also discuss what happens far from the transition line.

Fig.(\ref{compS1S2S3})-b,d show $\theta_2$ at the points  $(\aaa^{P_\text{i}}, \hh^{P_\text{i}}, \ww^{P_\text{i}}, \hh_2, \ww_2)$. For each $P_\text{i}$, we compute the contact angle $\theta_2$ using Eq.(\ref{en_CB}) and Eq.(\ref{en_W_s2}) and the minimization procedure for each pair $(\ww_2, \hh_2)$, with $\ww_2 \in (0, \ww^{P_\text{i}})$ and $\hh_2 \in (0, \hh^{P_\text{i}})$.
 
We now seek the {\it optimal surface 3}, which is the surface of type 3 that maximizes the  $\ttc$ compared to the surface of types 1 and 2. To find the optimal surface 3, we use the same method applied in the previous section to select the optimal surface 2. We recall  the procedure here, applying it for the surface of type 3. (i) For each set of parameters $(\aaa^{P_\text{i}}, \hh^{P_\text{i}}, \ww^{P_\text{i}}, \hh_2, \ww_2)$ of the surface of type 2, we vary the parameters of the surface of type 3: $\hh_3 \in (0, \hh_2)$  and $\ww_3 \in (0, (\ww-\ww_2)/2)$.  (ii) For each set of parameters $(\aaa^{P_\text{i}}, \hh^{P_\text{i}}, \ww^{P_\text{i}}, \hh_2, \ww_2, \hh_3, \ww_3)$, we minimize Eq.(\ref{en_CB}) and Eq.(\ref{en_W_s3}) and find the contact angle $\theta_3$ that minimizes the global energy of the droplet on this surface 3. (iii) After spanning all the possible geometries of the surface 3, we find $\theta_3^{\mm}(\theta_2)$, which is defined as the angle that maximizes the difference between $\theta_3$ and $\theta_2$. We also find $\theta_3^{\mm}(\theta_1)$, the angle that maximizes the difference between $\theta_3$ and $\theta_1$, but is it not shown here because the diagram of  $\theta_3^{\mm}(\theta_1)$ is similar to the diagram of  $\theta_3^{\mm}(\theta_2)$ for the points chosen.

The diagrams of Fig.(\ref{compS1S2S3})-b,c,d,e allow us to investigate, for any point  $(\aaa^{P_\text{i}}, \hh^{P_\text{i}}, \ww^{P_\text{i}}, \hh_2, \ww_2)$, the relation between the optimal surface of type 3 and the other surfaces with the same base $(\aaa^{P_\text{i}}, \hh^{P_\text{i}}, \ww^{P_\text{i}})$ but different types of reentrances.

The diagram of Fig.(\ref{compS1S2S3})-b shows $\theta_{2}$ of the point $P_1$ indicated in Fig.(\ref{compS1S2S3})-a.
The whole diagram presents $\theta_{2} \leqslant \theta_{1}$, meaning that the contact angle is never bigger in the surface 2 than it would be in the surface of type  1. This remains true for any point $P_\text{i}$ inside of the CB phase for the surface 1, which confirms the conclusion of the previous section: if the droplet were in a CB state in the surface of type 1, its contact angle keeps a high value when placed on a surface of type 2.
To understand if there is any gain in using a double reentrance, we choose  
two typical points of this diagram, identified as $\alpha$ and $\beta$.
For the $\alpha$ point, Fig.(\ref{compS1S2S3})-c shows that $\theta_3^{\mm} >\theta_{2}$, indicating that the optimal surface of type 3 enhances $\ttc$ compared to the surface of type 2, but
$\theta_3^{\mm} \approx \theta_{1}$.
For the $\beta$ point we observe that all diagrams have the same color, indicating that there is no significant difference in the $\ttc$ for all surfaces.
Both  situations are illustrated on the right of the diagrams where we also indicate the geometric parameters of the surfaces for the points $\alpha$ and $\beta$. The inequalities indicate the relation between the contact angles in different surfaces.

The situation is different when one considers a point $P_2$ that is in the W region in the pillared surface, shown in Fig.(\ref{compS1S2S3})-a.
In this case, the use of a double reentrant surface can enhance  significantly the contact angle.  
In the case of the $\gamma$ point, we observe in Fig.(\ref{compS1S2S3})-d that $\theta_2 < \theta_1$ but  $\theta^{\mm}_3 - \theta_1 \approx 15^{\circ}$, generating a relation expressed in the schema on the right.
Finally, the $\delta$ point is in a region where the differences between the surfaces are smoothed when compared with the $\gamma$ point.
Fig.(\ref{compS1S2S3})-d shows that  $\theta_2  \approx \theta_1$, but  Fig.(\ref{compS1S2S3})-e shows $\theta^{\mm}_3-\theta_2 \approx 10^{\circ}$. 
This situation is shown in the right of the diagram. 

To close this section,  we comment on the wetting behavior of surfaces with the geometries given by the points $P_\text{i}$ of the diagram Fig.(\ref{compS1S2S3})-a that are far from the phase transition line.
If $P_\text{i}$ is in the CB phase, the behavior observed in the point $\alpha$ disappears and the situation explained in the point $\beta$ is dominant.  If $P_\text{i}$ is far from the transition line but in the W phase, the dominant behavior is the one discussed in $\delta$ point; the situation shown in the $\gamma$ disapears.

\subsection*{Qualitative comparison with experiments}
In this section we compare the results of our model with some recent experiments that use reentrant surfaces \cite{Liu2014, chen2015, Ramos2016}. We discuss some features that can be qualitatively described by the model and the limitations of the global energy  approach.

\subsubsection*{Contact angle of a droplet as function of $\tty$}

In the reference \cite{Liu2014} Liu and Kim have shown that while the pillared surfaces could not sustain the super-repellent character for liquids with surface tension below $\sGL \approx 50 mN/m$, the introduction of a simple reentrance in the surface extend its super-repellent behavior up to liquids with $\sGL\approx 20 mN/m$ and the addition of a double reentrance in the surface allowed it to become super-repellent  even for liquids with $\sGL\approx 10 mN/m$.

To understand to which extent our model is able to describe the results reported in \cite{Liu2014} and better explore the wetting behavior of the  configurations encountered before for different types of liquids, we select some specific geometries that produce the five possible wetting relations shown in Table(\ref{tabComparaE2}) as a function of $\tty$. 
Despite the fact that the Young angle is a result of the interaction between the liquid placed on a flat surface prepared with a given chemistry, we remind that in our model $\tty$ is the only parameter related to the type of liquid. 
We assume that different values of $\tty$ mimics liquids with different values of surface tension $\sGL$ when placed on a surface with the same chemistry.  In other words, $\tty$ is an effective way of changing the type of liquid.

\begin{figure}[H]
\centering
\includegraphics[width = 1\columnwidth]{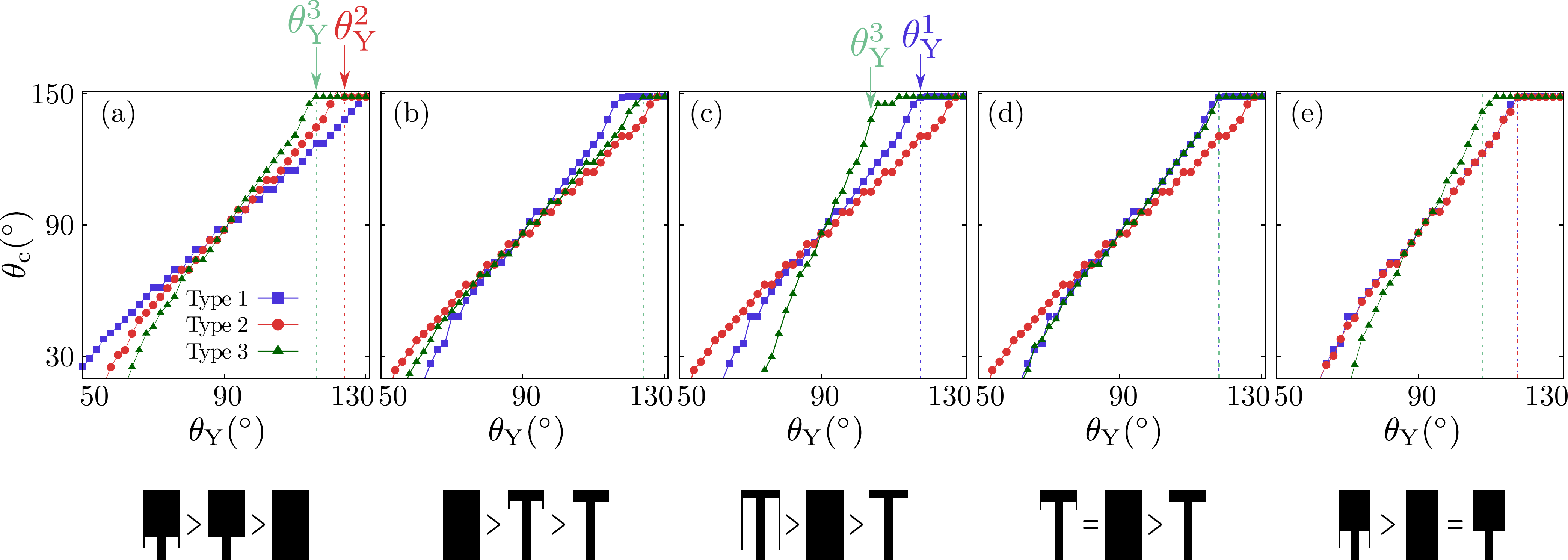}
\caption{Contact angle $\ttc$  as a function of $\tty$. Different colors correspond to $\ttc$ for different types of surfaces, as indicated in the legend box. For all figures $\ww=\hh=\aaa=50\,\mu m$ and  $\ww_2=\ww_3=1\,\mu m$, which results in a solid fraction $\Phi_S=\ww^2/(\ww+\aaa)^2 \approx$ $0.25$ and $\hh_2^{\ast}=25.5\,\mu m$. The relation between geometric parameters of the surfaces are shown in the schema below each figure and the comparative symbols refer to  $\ttc$ on each surface for the case $\tty>90^{\circ}$. The geometric parameters and the respective condition in parenthesis are given by: {\bf(a)} $\hh_2=10\,\mu m$, $\hh_3=8\,\mu m$ ($\hh_2 < \hh_2^{\ast}$). 
{\bf(b)} $\hh_2=49\,\mu m$, $\hh_3=5\,\mu m$ ($\hh_2 > \hh_2^{\ast}$ ~and~ $\hh_3<\hh_3^*$). {\bf(c)} $\hh_2=49\,\mu m$, $\hh_3=48\,\mu m$ ($\hh_2 > \hh_2^{\ast}$ ~and~ $\hh_3>\hh_3^*$).  {\bf(d)} $\hh_2=49\,\mu m$, $\hh_3=11.6\mu m$ (~$\hh_2 > \hh_2^{\ast}$ ~and~ $\hh_3=\hh_3^*$~). For the cases b,c and d: $\hh_3^*=11.6\,\mu m$.~ {\bf(e)}  $\hh_2=25.5\,\mu m$, $\hh_3=10\,\mu m$ ($\hh_2 = \hh_2^{\ast}$). $\tty^{i}$ is the value of $\tty$ which the droplet transit between states for the surface of type $i$.}
\label{CA_vs_ty}
\end{figure}

Fig.(\ref{CA_vs_ty}) summarizes all possible relations between the wetting behavior of the three types of surfaces. It shows the  $\ttc$ of a droplet  in the thermodynamic stable state on each surface as a function of $\tty$. The values of $\ttc$ were obtained by fixing each geometry and, for each value of $\tty$, we applied the minimization procedure.

Considering for example Fig.(\ref{CA_vs_ty})-c, we note that for very high value of $\tty$,  for surfaces of type 1 and type 3  the thermodynamic state of a droplet is the CB state. For the range of $\tty$ presented in the figure, a  thermodynamic state  of a droplet placed on the surface of type 2  would be W.
When $\tty$ decreases, $\ttc$ on a surface of type 1 would make a transition for the W state at the $\tty=\tty^{1}\approx 120 ^{\circ}$, while the thermodynamic state of droplet placed on surfaces of type 3 would keep the CB state up to $\tty=\tty^{3}\approx 104 ^{\circ}$. 
We stress that the qualitative behavior shown in Fig.(\ref{CA_vs_ty}) is robust in the sense that it would happen for different values of solid fraction $\Phi_S$. However, depending on $\Phi_S$, the same behavior would be observed for a different range of geoemtric parameters and $\tty$.

Besides the rich variety of the wetting behavior presented by all these relations, the model is not able to describe the experimental result shown in reference \cite{Liu2014}. 
An important limitation of the model, based on the global energy minimization, is that it does not describe the super-repellent behavior for surfaces with $\tty<90^{\circ}$, as it was theoretically discussed for example in reference \cite{Patankar2009} and anticipated by us in a previous section.
Moreover, the relation between the contact angles of different surfaces found in \cite{Liu2014} is the condition shown in Fig.(\ref{CA_vs_ty})-a, for which $\tty^{3}> \tty^{2}>\tty^{1}$. However, in our model the geometric conditions of the surfaces that produces such relation is very different from the configurations used in \cite{Liu2014}: while in \cite{Liu2014} the surface of type 2 has high value of $\hh_2$ and the surface o type 3 has a small $\hh_3$, in our case the value of $\hh_2$ is small and $\hh_3$ is relatively big as shown in the schema below the figure and written in the caption of the figure.
We will show in the next section that if the initial state of the droplet in the simulations is a CB state, it stays in this repellent behavior even though the thermodynamics predicts that the final state should be  W. It suggests that there is a barrier to transit from CB to W state that leads to a metastability of the CB state and offers an explanation for the disagreement between the model and the experiment. 

\subsubsection*{Evaporation on the reentrant surfaces}

In references \cite{chen2015,Ramos2016} the authors report evaporation experiments  of the droplet on surfaces with reentrances. 
In \cite{Ramos2016} they study the influence of the solid-liquid fraction of the surfaces and the temperature of the substrate on the evaporation of the droplet placed on a superhydrophobic surface with reentrant micropillars. 
The work \cite{chen2015}  focus on the difference of the evaporation dynamics for liquids with low and high surface tensions placed on the surfaces with reentrant mushroom structures on copper substrates.

\begin{figure}[H]
\centering
\includegraphics[width = 1\columnwidth]{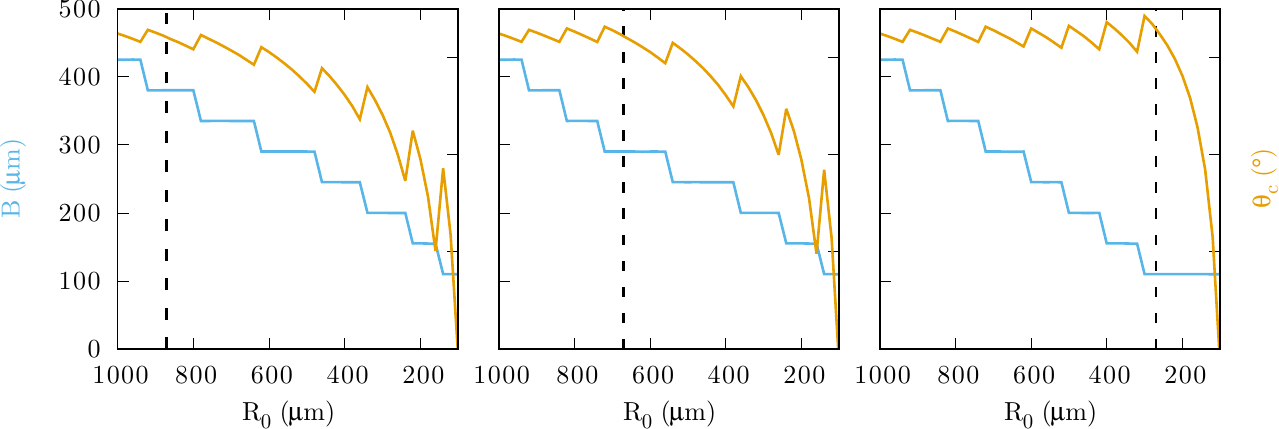}
\caption{Base radius $B$ and contact angle $\ttc$ as a function of initial radius $R_0$. Solid lines correspond to the thermodynamic stable values for the contact angle. Vertical dashed line marks the passage from CB to W state when the radius reduces. The geometrical parameters are $\ww = 40\mu$m, $\aaa=50\mu$m, $\hh=50\mu$m, $\ww_2=10\mu$m, $\hh_2=20\mu$m, $\ww_3=2\mu$m and $\hh_3=10\mu$m.}
\label{figevap}
\end{figure}

To compare our model with these experiments, we mimic the evaporation dynamics by changing the initial volume of the droplet.
We note that Eq.(\ref{en_CB})-Eq.(\ref{en_W_s3}) are modified when the droplet's volume is reduced, since the terms $N^{\rm s}_{\bf (i)}$ and $S^{\rm s}_{\bf (i)}$  in these equations depend on the droplet radius $R$. 
Fig.(\ref{figevap}) shows the contact angle  $\theta_C$ and the basis radius $B$ as a function of the droplet's initial radius $R_0$ for specific geometries of the three different types of surfaces. Vertical lines indicate the passage from the CB to W state when reducing $R_0$. 

Our model do not have quantitative agreement with the experiments, but it is able to describe qualitatively some features reported in the experiments \cite{chen2015,Ramos2016}: i) we do observe a transition from CB to W state when the volume of the droplet reduces, ii) there is a ``staircase'' behavior of $\theta_C$ and $B$ and iii)  the surface of type 3 is able to sustain a high value of the contact angle for smaller values of volumes for this particular geometry of the surface we chose. Features i) and ii) have already been reported for pillared surfaces experimentally~\cite{Ramos2015},  in simulations \cite{Fernandes2015, Shahraz2013} and  more recently for the reentrant surfaces \cite{chen2015,Ramos2016}. The staircase behavior in our case is due to the fact that the energy is minimized  subject to the constraint that the contact line is pinned. In experimental systems this behavior was classified as a complex mode characterized by a series of stick-slip events \cite{Ramos2015, Ramos2016}.

\section*{Numerical experiments}

The theoretical model discussed in this work takes into account the global energy of the droplet and allow to predict its geometrical properties at the stable thermodynamic state.
It is known, however, that final state of the droplet may change if it is carefully deposited or thrown on the substrate \cite{Quere2005}. This exemplifies  that  in some situations the droplet gets trapped in a metastable state and do not reach its equilibrium state; to transit from one state to another it is then necessary to overcome an energy barrier \cite{Giacomello2012,Zhang2015,TutejaPNAS2008,giacomello2016perpetual}.

In this section we perform numerical simulations using Monte Carlo method of the cellular Potts model to better understand the dependency of the initial wetting state of the droplet on its final state. 
The details of the model and parameters used in the simulations for pillared surfaces are explained in the reference \cite{Fernandes2015} and in the SI. In this work we change the geometry of the substrate and perform simulations for different geometric parameters of the reentrant surfaces. The analyses shown here are for $\tty = 114^{\circ}$. 

\begin{figure}
\centering
\includegraphics[width=1\linewidth]{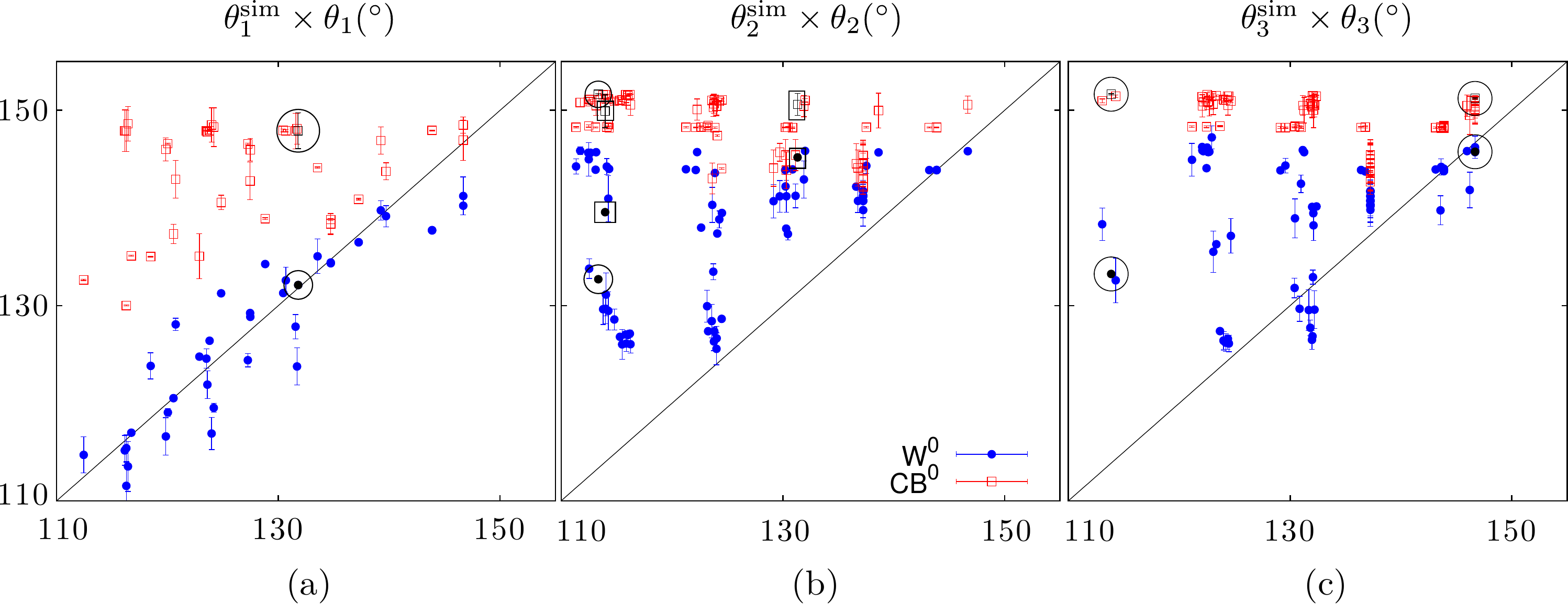}
\caption{Scatter plot of stationary contact angles as a function of the theoretical values for {\bf (a)} surface of type 1, {\bf (b)} type 2 and {\bf (c)} type 3. The (blue) circles correspond to simulations starting in the W$^0$ configuration while the (red) squares in the CB$^0$ state. The black line is the expected relation of  equality between simulations and theory. Points are averages over $5$ simulation runs for $R_0=100\,\mu m$ and for various values of geometric sets. The geometric parameters $\ww=10\,\mu m$,   $\ww_2=2\,\mu m$ and  $\ww_3=1\,\mu m$ are fixed for all points of the three surfaces.  The big black circles indicate the geometric surfaces shown in the cross section of the Fig.(\ref{snap1}) and big black squares correspond to the geometries shown in the Fig.(\ref{snap2}).}
\label{scatter}
\end{figure}

\begin{figure}
\includegraphics[width = 1\columnwidth]{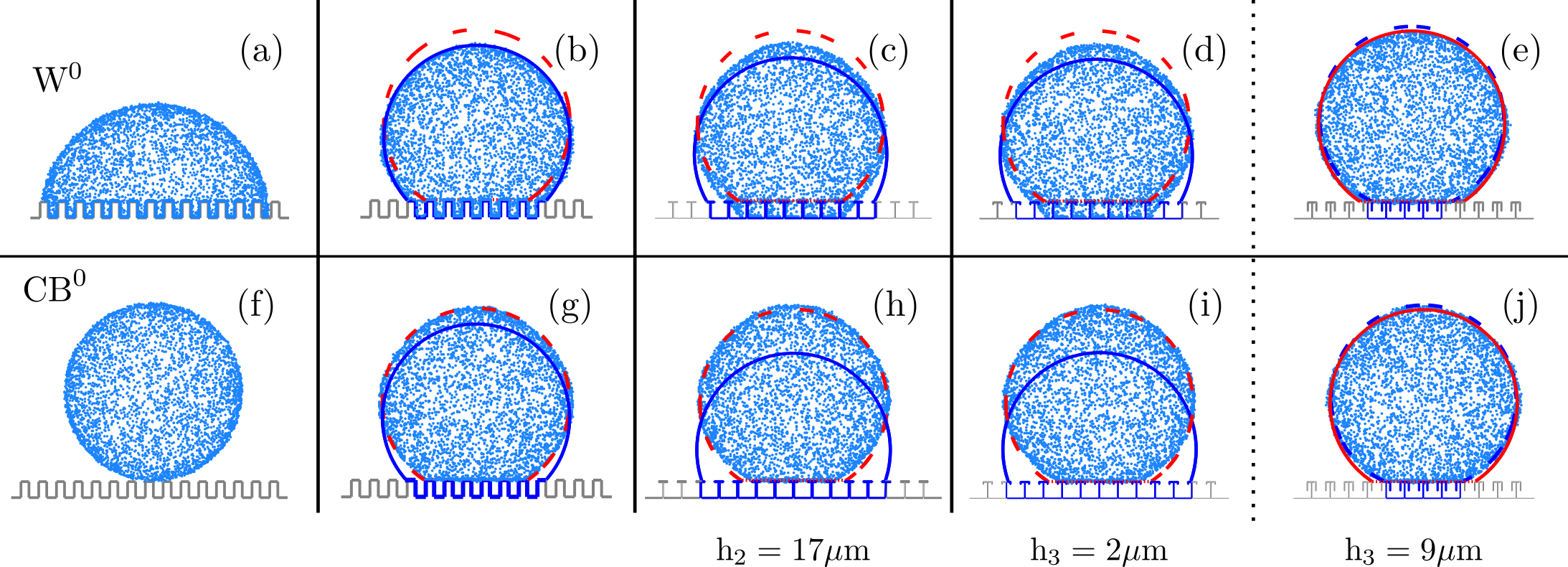}
\caption{Cross section of the droplet configuration in the final state of the Monte Carlo simulation, starting from the W$^0$ configuration (above) and from the  CB$^0$ configuration (below). The  blue line represents the cross section for the minimum energy W configuration and red line represents the cross section for the minimum energy CB configuration. Solid line identifies  when the solution is a global minimum. The snapshots correspond to droplets with $R_0=100\,\mu m$ placed on a surface with fixed interpillar distance and pillar width and pillar height ($\aaa=10\,\mu m$, $\ww=10\,\mu m$, $\hh=18\,\mu m$, $\ww_2 = 2\,\mu m$ and $\ww_3 = 1\,\mu m$). Other geometric parameters that defines each type of surface are given by: (c),(h) surface of type 2 with ($\hh_2=17 \mu $m), (d),(i) surface of type 3 with ($\hh_2=17 \mu$m, $\hh_3=2 \mu$m) and (e),(j) surface of type 3 with ($\hh_2=17 \mu$m, $\hh_3=9 \mu$m). It is interesting to note that when the initial state is CB$^0$ the final state of the droplet coincides with the minimum CB configuration, which is a local minimum.}
\label{snap1}
\end{figure}

The simulations do not allow to measure the size of the energy barrier, but it allow us to discuss how difficult is to reach the thermodynamic wetting state predicted for different geometries when the initial wetting state changes. 
To test this dependence on the initial wetting state, the droplet is initialized in two different wetting regimes. All the simulated contact angles for each initial state are summarized in the Fig.(\ref{scatter}), which clearly shows that the final contact angles are different when initializing in different wetting states. The two wetting states are generated as follows.
One possible wetting initial state is exemplified in Fig.(\ref{snap1})-a.
It is created using a hemisphere with the initial volume $V_0 \approx V_T =4/3\,\pi R_0^3$. We refer to this state as an initial Wenzel state, W$^{0}$.
The second possible wetting state is exemplified in Fig.(\ref{snap1})-f.
In this case a droplet with the same initial volume $V_0$ as in the W$^{0}$ state is placed slightly above the surface and allowed to relax under the influence of gravity. 
Because the droplet is not filling the surface, we refer to this as an initial Cassie-Baxter state, identified as CB$^{0}$. 
Due to numeric resources limitations and the need to span a big range of parameters, we simulate a droplet of radius $R_0 = 100 ~ \mu m$ which is  much smaller than the size of the droplet considered in the previous sections.
The total run of a simulation is at most $5\times10^5$ MCS  (Monte Carlo Steps, better explained in the SI) for each geometry and the last $1\times10^5$ MCS are used to measure observables of interest. Even with this long transient time,  for some initial conditions the system does not reach the thermodynamically stable state and becomes trapped in a metastable state. At least $5$ different initial conditions are used for each set of simulation parameters.

Fig.(\ref{scatter})-a,b,c show scatter plots to compare quantitatively the contact angle of the droplet obtained theoretically and in simulations for the three different surfaces. The horizontal axis presents the theoretical values, $\theta_i$ with $i=1,2,3$ and the vertical axis show the results from simulations $\theta_i^{\textrm{sim}}$. 
Black line represent points for which $\theta_i^{\textrm{sim}}=\theta_i$. Then, closer are the points to this line, better is the agreement between theory and simulation.
Each point on the scatter plot are averages over 5 simulation runs for a given set of geometric parameters and the error bars correspond to the standard deviation of the average. All the points simulated are shown in Fig.(S2), together with its predicted thermodynamic state and $\theta_i^{\textrm{sim}}$. 

Let us first compare the theoretical predictions with the results of simulations for the surface of type 1, shown in Fig.(\ref{scatter})-a.  When the initial state is W$^0$ as shown in Fig.(\ref{snap1})-a,  $\theta_1^{\textrm{sim}}$ present a good agreement with the theory, because the simulations are better able to explore the phase space and to make the transition to the CB state. However, when the initial state is CB$^0$, Fig.(\ref{snap1})-f, the agreement between simulations and theory is good only in the region where the thermodynamically stable state is the CB one. This means that, when the initial state is W$^0$  and the thermodynamic state is CB, the droplet is able to change its state (during a simulation run, all samples reach the predicted state). On the other hand, when the theoretically predicted thermodynamically stable state is W and the droplet is initialized in the CB$^0$ wetting state,  the droplet is generally not able to overcome the barrier between the states and becomes trapped in the metastable CB state.  
The same behavior is reported in the reference \cite{Fernandes2015} for smaller droplets.
This metastability of the Cassie-Baxter state is in agreement with the observation made in experiments~\cite{McHale2005} and (almost) 2D systems simulated by means of molecular dynamics of nanodroplets~\cite{Shahraz2012,Shahraz2013} and it is consistent with the existence of a high energy barrier between the  thermodynamical states: as $h$ gets higher, it becomes increasingly more difficult for the system to go from the CB state to the W state. 

The metastability of the Cassie-Baxter state observed for the pillared surface is also encountered for surfaces with simple and double reentrances, Fig.(\ref{scatter})-b and Fig.(\ref{scatter})-c respectively.
Moreover, for these reentrant surfaces, the agreement between theory and simulations happens for very few cases even when the initial wetting state is W$^0$: most of the simulated angles $\theta^{\textrm{sim}}$ are higher than $\theta_i$ as it is shown by the points above the line $\theta_i^{\textrm{sim}}=\theta_i$.
We analyzed the geometries of the points that are closer to the line $\theta_i^{\textrm{sim}}=\theta_i$  to understand why the agreement is better for some geometries. Although it was not possible to extract a general rule for that, we identified that these points are more likely to correspond to geometries such that in the pillared surface the parameters $(\aaa, \hh, \ww)$ corresponded to the region of W state. 
In other words, if the pillared surface had a repellent behavior (CB wetting thermodynamic state), adding reentrances do not have the influence predicted by the model.

\begin{figure}
\includegraphics[width = 1\columnwidth]{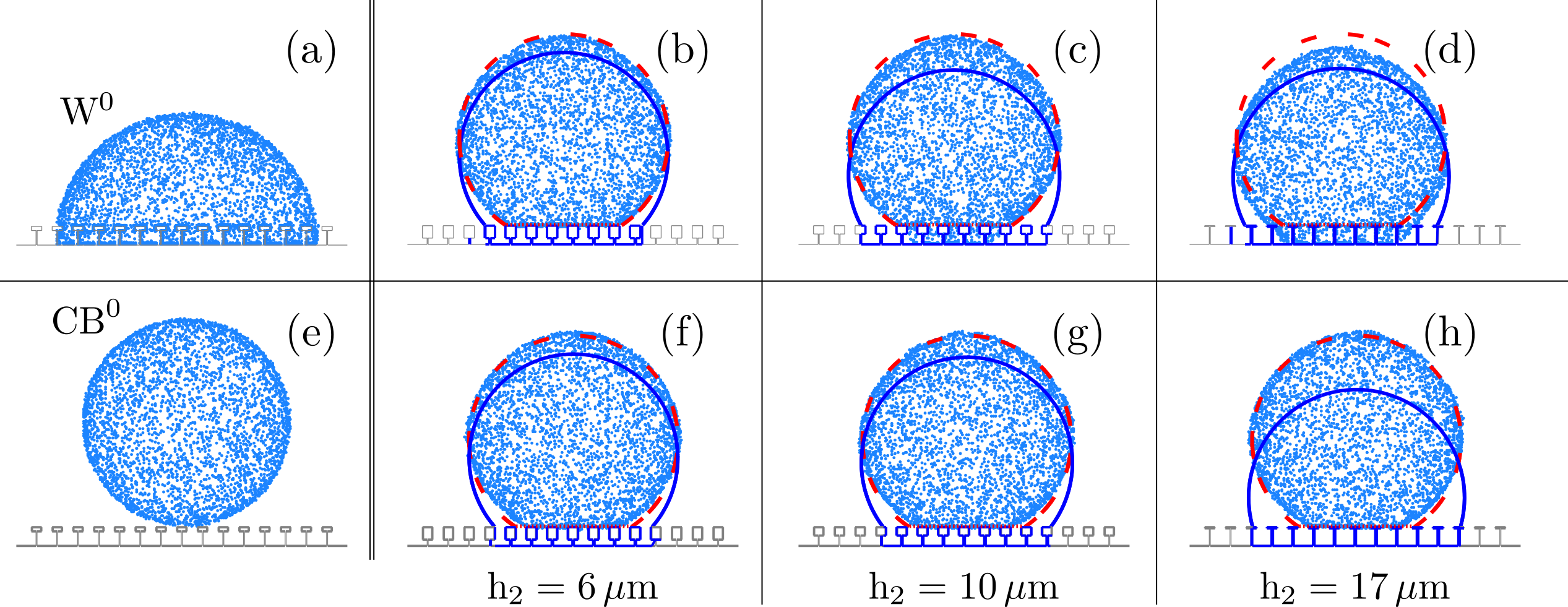}
\caption{The snapshots correspond to droplets with $R_0=100\,\mu m$ placed on a surface of type 2 with fixed interpillar distance and pillar width and pillar height ($\aaa=10\,\mu m$, $\ww=10\,\mu m$, $\hh=18\,\mu m$ and $\ww_2 = 2\,\mu m$) and varying $\hh_2$. The notation for different cross sections is the same as in the Fig.(\ref{snap1}).}
\label{snap2}
\end{figure}

Fig.(\ref{snap1}) and Fig.(\ref{snap2}) show cross sections of final droplet configurations for different surfaces and different initial wetting conditions obtained from MC  simulations. To compare with the continuous model, it is shown together the resultant cross sections that correspond to both $\Delta E^{\bf {W}}_{\bf min}$ (blue line) and $\Delta E^{\bf {CB}}_{\bf min}$ (red line). Note that one of these two states is the global minimum and it is identified by the continuous line, while the dashed line represents the local minimum. The initial wetting state, W$^0$ or CB$^0$, is indicated by the first image of each line.  These snapshots are useful to visualize the solutions and illustrate some of the observations we draw based on our simulations:
i) $\theta_2^{\textrm{sim}}$ > $\theta_1^{\textrm{sim}}$ (only one exception is observed)  irrespectively of the theoretical predicted relation between the theoretical angles. It happens for both initial wetting states W$^0$ or CB$^0$, although this effect is more important when the initial wetting state is W$^0$. ii)  $\theta_3^{\textrm{sim}} \approx \theta_2^{\textrm{sim}}$ for most of the geometries. Cases where $\theta_3^{\textrm{sim}} >  \theta_2^{\textrm{sim}}$ are more likely to happen for geometries with big values of $\hh_3$. An example can be visualized in Fig.(\ref{snap1}):  when the initial state is W$^0$, $\theta_3^{\textrm{sim}}> \theta_2^{\textrm{sim}}> \theta_1^{\textrm{sim}}$. Moreover, the contact angle $\theta_3^{\textrm{sim}}$ increases when $\hh_3$ increases, Fig.(\ref{snap1})-d,e.
iii) For the surface of type 2, if parameters $(\aaa, \hh, \ww, \ww_2)$ are fixed  and  $\hh_2$ increases, the $\theta_2^{\textrm{sim}}$ decreases. An example of the role of $\hh_2$ on the final state of the droplet is shown in the Fig.(\ref{snap2}).
It is interesting to observe that in both examples, Fig.(\ref{snap1}) and Fig.(\ref{snap2}), when the initial wetting state is CB$^0$, the final state of the droplet do not reach the global minimum, but it coincides with the minimum CB state, which is a local minimum. 

\section*{Summary and Conclusions}

In this work we extend a simple model previously applied to pillared surfaces \cite{Fernandes2015} for reentrant surfaces of the type shown in Fig.(\ref{typesSurface}). The model is developed to understand the wetting state of a three-dimensional droplet when  placed on a pillared and reentrant surfaces based on the analysis of the total interfacial energies associated with the two possible wetting states, W and CB.

From the analysis of the equations of the model in the limit where the droplet volume is big compared to the roughness of the surface, we are able to derive analytically the geometric relations between the energy of the droplets on each type of surface that would enhance the CB state. 
These analysis show that the wetting behavior of the three surfaces are governed by some non trivial relations between the height $\hh_2, \hh_3$ and the width $\ww_2, \ww_3$ of the reentrances, which are summarized in the Table (\ref{tabComparaE2}).
Due to the minimization procedure we  find the stable wetting state for each geometry and the corresponding contact angle $\ttc$ of the droplet in this state. 
We then span the geometric parameters for each type of surface and, by comparing the thermodynamic contact angle that the  droplet would have if  placed on these surfaces, we find the type of geometries that most enhances  the apparent $\ttc$ of the droplet.  Both the theoretical analysis and the minimization process allow us (i) to quantify the differences in the $\ttc$ for all possible relations between the 3 surfaces as a function of the type of liquid,  as summarized in Fig.(\ref{CA_vs_ty}) and (ii) to find some geometries that enhances the thermodynamic contact angle and keeps the super-repellent behavior for liquids with smaller surface tension as  for the example shown in Fig.(\ref{CA_vs_ty})-c.

The global energy approach is known to have limitations~\cite{gao2007wenzel, Marmur_CWright, gao2009attempt, Nosonovsky2007,Enright2012,Patankar2009}  and success, describing for instance qualitatively the dependency of the wetting state on the initial volume of the droplet~\cite{Tsai2010, Xu2012,Fernandes2015}. In the context of the reentrant surfaces, the thermodynamic approach fails in describing the super-repellent behavior of surfaces build from materials for which $\tty < 90^{\circ}$, as it has been shown to be possible  experimentally for different groups \cite{TutejaSicence2007, TutejaPNAS2008,Liu2014}. 
Recent MD simulations have shown that simple reentrant surfaces does increase the barrier to pass from the CB to W state even for the case where $\tty< 90^{\circ}$, which can explain why even though W is the thermodynamic state, dynamic barriers make it difficult to reach the most stable state promoting a metastability of the repellent behavior \cite{Quere2008,TutejaSicence2007,joly2009}.
To address this important debate, we implemented Monte Carlo simulations. Although our simulations do not allow to measure the size of the barrier between the repellent and wet states, we can observe the difficulty to bypass the barrier between the two wetting states by changing its initial wetting state. For all types of surfaces studied in this work and $\tty>90^{\circ}$, we observed that, once initialized in CB$^0$ state, the droplet gets trapped in the local minimum that corresponds to the minimum of the CB state predicted by the theoretical model. When the droplet is initialized in the W$^0$ state, the agreement between theory and simulations is good in the case of pillared surfaces, but for reentrant surfaces we observe that the final contact angle of the droplet in the simulations is higher than predicted by the model for most of the geometries that we considered. 

It would be useful to quantify the size of this barrier as a function of the geometric parameters of the reentrant surfaces.
A possible way to do a quantitative estimation of the barrier using Monte Carlo simulations is to implement for example a method like "umbrella sampling" \cite{Frenkel2002167}.
Other improvement of our model would be to take into account the curvature of the hanging liquid-air interface that in our model is considered flat \cite{Quere_EPL2008}.  
It would also be interesting to take into account in the case of the reentrant surfaces the role of pressure that the liquid volume exerts to impale the surface \cite{Patankar_Langmuir2010} and some analysis of the barrier for liquid impalement \cite{Zhang2015}.

{\bf Supporting Information (SI)}: it is described the algorithm  to find the thermodynamic wetting state for a droplet placed on different types of surfaces considered in this work, the equations of the volume of the droplet in each type of surface, the details of the Potts Model used the simulations and a table with simulation results for all the geometric parameters considered.

\begin{acknowledgement}
The authors thank Heitor C. M. Fernandes, Mendeli H. Vaistein  for  discussions and for sharing with us the simulations codes  and Stella M. M. Ramos for nice discussions. We also acknowledge the use of the Computational Center of the New York University (NYU). M.S. thanks Rafael Barfknecht for helping with graphical softwares and technical comments.
\end{acknowledgement}

\bibliography{droplet}

\end{document}